\documentclass[twocolumn]{article}
\makeatletter
\let\ifaastex\@secondoftwo
\let\ifarxiv\@firstoftwo
\makeatother
\usepackage{ifxetex,ifluatex}
\ifxetex\else\ifluatex\else\usepackage[utf8]{inputenc}\fi\fi
\ifaastex{}{\usepackage{font-termes}}
\ifaastex{
  \usepackage[strict]{aaspatch}
  \usepackage{acro,booktabs,csquotes,hyperref,siunitx,tikz}
  \hypersetup{hidelinks}
}{
  \usepackage{authblk,booktabs,hyperref,tikz}
  \usepackage{basic,physics,astro,draft}
  \usepackage[mode=biblatex, hyperref]{apjbib}
  \ExecuteBibliographyOptions{embedlinks}
  \addbibresource{torus.bib}
}
\ifluatex\hypersetup{pdfencoding=auto}\fi

\newcommand*\citepaliasfirst[1]{%
  \citetext{\citealp{#1}\postnotedelim
  \bibstring{citedas}\addspace\citetalias{#1}}}
\defcitealias{2016ApJ...825...67C}{CK16}
\makeatletter
\newcommand*\Citetseealso[3]{%
  \Citeauthor{#1}\nameyeardelim\citetext{%
  \ifx\@empty#2\@empty\citeyear{#1}\else\citeyear{#1,#2}\fi
  \multicitedelim\citealp[see also][]{#3}}}
\makeatother
\makeatletter
\ifarxiv{\@ifpackageloaded{astro}{\let\ast@symit\textit}{}}{}
\makeatother
\input{torus.acr}
\newcommand*\xray{\texorpdfstring{X\protect\nobreakdash-ray}{X-ray}}
\newcommand*\xrays{\xray s}
\newcommand*\uvrmhd{\ac{UV}\protect\nobreakdash-\acp{RMHD}}
\newcommand*\irrmhd{\ac{IR}\protect\nobreakdash-\acp{RMHD}}
\newcommand*\titleuvrmhd{\texorpdfstring
  {\acs*{UV}\protect\nobreakdash-\acsp*{RMHD}}{UV-RMHD}}
\newcommand*\titleirrmhd{\texorpdfstring
  {\acs*{IR}\protect\nobreakdash-\acsp*{RMHD}}{IR-RMHD}}
\newcommand*\momsrc[1]{\vec S^\su m_\su{#1}}
\newcommand*\momsrcmag[1]{S^\su m_\su{#1}}
\ifarxiv
  {\newcommand*\ergsrc[1]{S^{\mkern-2mu\su e}_{\mkern-3.5mu\su{#1}}}}
  {\newcommand*\ergsrc[1]{S^\su e_\su{#1}}}
\ifaastex{}{}
\begin{document}

\title{Geometrically thick obscuration by radiation-driven outflow from
magnetized tori of active galactic nuclei}
\author{Chi-Ho Chan}
\affil{Racah Institute of Physics, Hebrew University of Jerusalem, Jerusalem
91904, Israel}
\affil{School of Physics and Astronomy, Tel Aviv University, Tel Aviv 69978,
Israel}
\author{Julian~H. Krolik}
\affil{Department of Physics and Astronomy, Johns Hopkins University,
Baltimore, MD 21218, USA}
\date{May 10, 2017}

\shorttitle{Radiation-driven obscuring AGN torus outflow}
\shortauthors{Chan \& Krolik}
\pdftitle{Geometrically thick obscuration by radiation-driven outflow from
  magnetized tori of active galactic nuclei}
\pdfauthors{Chi-Ho Chan, Julian H. Krolik}

\begin{abstract}
\ifaastex\raggedright{}
Near-Eddington radiation from \acp{AGN} has significant dynamical influence on
the surrounding dusty gas, plausibly furnishing \acp{AGN} with geometrically
thick obscuration. We investigate this paradigm with \aclp{RMHD} simulations.
The simulations solve the \aclp{MHD} equations simultaneously with the \ac{IR}
and \ac{UV} \ac{RT} equations; no approximate closure is used for \ac{RT}. We
find that our torus, when given a suitable sub-Keplerian angular momentum
profile, spontaneously evolves toward a state in which its opening angle,
density distribution, and flow pattern change only slowly. This
\textquote{steady} state lasts for as long as there is gas resupply toward the
inner edge. The torus is best described as a mid-plane inflow and a
high-latitude outflow. The outflow is launched from the torus inner edge by
\ac{UV} radiation and expands in solid angle as it ascends; \ac{IR} radiation
continues to drive the wide-angle outflow outside the central hole. The dusty
outflow obscures the central source in soft \xrays, the \ac{IR}, and the
\ac{UV} over three quarters of solid angle, and each decade in column density
covers roughly equal solid angle around the central source; these obscuration
properties are similar to what observations imply.
\end{abstract}
\acresetall\acuseshort

\section{Introduction}

Observations concur on the existence of geometrically and optically thick
toroidal obscuration in \acp{AGN} \citep[e.g.,][]{1985ApJ...297..621A,
1990ApJ...355..456M}; in fact, the torus is an essential ingredient in the
unification of type\nobreakdash-1 and type\nobreakdash-2 \acp{AGN}
\citep[e.g.,][]{1989ApJ...336..606B, 1993ARA&A..31..473A, 1995PASP..107..803U}.
We observe several type\nobreakdash-2 \acp{AGN} for every type\nobreakdash-1
\ac{AGN}, but the exact ratio remains contentious
\citep[e.g.,][]{2008A&A...490..905H, 2010ApJ...714..561L}.

Despite consensus that geometrically thick obscuration exists, there is little
agreement on how the torus remains inflated in the deep gravity well of the
\ac{SMBH} while being cool enough to hold on to its dust. Attempts at resolving
this problem have invoked a variety of mechanisms: warped disks
\citep[e.g.,][]{1989tad..conf..457P, 1989ApJ...347...29S}; clumping
\citep{1988ApJ...329..702K}; magnetic support, either static
\citep{1998A&A...338..856L} or in winds \citep[e.g.,][]{1994ApJ...434..446K,
2006ApJ...648L.101E}; starbursts \citep[e.g.,][]{2009MNRAS.393..759S,
2009ApJ...702...63W}; radiation pressure
\citetext{\citealp[e.g.,][]{1992ApJ...399L..23P, 2008ApJ...679.1018S,
2015ApJ...812...82W, 2016ApJ...819..115D}\multicitedelim\citealp[but
see][]{2016MNRAS.460..980N}}; and combinations thereof
\citep{2012ApJ...749...32K, 2016ApJ...828L..19W}. Unfortunately, as we
discussed in an earlier article \citepaliasfirst{2016ApJ...825...67C}, none of
these proposals provides a complete explanation.

Whatever dynamical processes operate, they must explain the structure of these
parsec-scale dusty regions as detected by \ac{MIR} interferometry in nearby
\acp{AGN}. Some \acp{AGN} reveal two components, one elongated in the
equatorial direction and the other in the polar direction
\citep[e.g.,][]{2004Natur.429...47J, 2009MNRAS.394.1325R, 2014A&A...563A..82T}.
The equatorial component is understood as the torus inner edge
\citep[e.g.,][]{2004Natur.429...47J}, while the polar component could come from
optically thin dust in the polar regions under direct illumination by the
central source \citep{2012ApJ...755..149H}. Polar emission is also seen in
\ac{MIR} imaging on \SI{\gtrsim100}{\parsec} scales
\citep[e.g.,][]{1993ApJ...409L...5B}. Circinus is a particularly striking
example: \Ac{MIR} emissions on \SI{\sim1}{\parsec} and \SI{\sim100}{\parsec}
are aligned with each other \citep{2016ApJ...822..109A}, and also with the edge
of the ionization cone \citep{2005ApJ...618L..17P, 2016ApJ...822..109A}. This
observation suggests that dust is concentrated along the edge of the cone
\citep{2012ApJ...755..149H, 2016ApJ...822..109A}, which could be the case if
radiation pressure on dust drives a dusty outflow from the inner surface of the
torus \citep{2012ApJ...755..149H}.

Several authors have tried to explain dusty outflows.
\Citet{2012ApJ...759...36R} performed Monte Carlo \ac{RT} on a static
distribution of dusty gas and found that radiative acceleration can overcome
gravity. \Citet{2012ApJ...749...32K} constructed analytic solutions of a
magnetocentrifugal wind where radiation pressure on dust provides additional
driving. Detailed \acp{RHD} simulations have also been performed. The
simulations by \citet{2016ApJ...828L..19W} considered momentum deposition by
\ac{UV} radiation on dust, but ignored momentum transfer between \ac{IR}
radiation and dust. \Citet{2016ApJ...819..115D} took the opposite approach by
including only momentum coupling between \ac{IR} radiation and dust.
\citetalias{2016ApJ...825...67C} was the first attempt to capture the effects
of both \ac{IR} and \ac{UV} radiative support in one single simulation. Despite
their significant differences, all these studies agree that, for a torus
orbiting a central object of mass $M\sim\SI{e7}{\solarmass}$ radiating at an
Eddington ratio $L/L_\su E\sim0.1$, the typical mass loss rate due to
radiation-driven dusty outflow is $\SI{0.1}{\solarmass\per\year}\lesssim\dot
M\lesssim\SI{1}{\solarmass\per\year}$. This is not a feeble outflow: A torus of
Thomson optical depth $\tau_\su T$ located just outside the dust sublimation
surface can be depleted by mass loss in
\begin{equation}
\num{\lesssim14}\,\tau_\su T
  \biggl(\frac M{\SI{e7}{\solarmass}}\biggr)^{3/4}
  \biggl(\frac{L/L_\su E}{0.1}\biggr)^{1/4}
  \biggl(\frac{\dot M}{\SI{0.1}{\solarmass\per\year}}\biggr)^{-1}\,
  \mtext{orbits.}
\end{equation}
This means a torus cannot remain in a steady state for many orbits unless there
is a constant resupply of gas from galactic scales through the torus down to
the inner edge \citep[see also][]{1988ApJ...329..702K, 2012ApJ...759...36R,
2014MNRAS.445.3878S, 2015ApJ...812...82W}.


\Ac{MHD} stresses generated by the \ac{MRI} are widely accepted as the
mechanism for angular momentum transport and mass accretion in geometrically
thin disks \citep[e.g.,][]{1991ApJ...376..214B, 1991ApJ...376..223H}; such
stresses conceivably regulate accretion in geometrically thick disks as well.
The \ac{MRI} grows in the ideal \acp{MHD} condition, which holds even when the
gas is weakly ionized \citep{1994ApJ...421..163B, 1996ApJ...457..355G}. Such
low levels of ionization can be sustained in the torus interior by \xrays{}
\citep{1995ApJ...447L..17N} since the energy produced by \acp{AGN} in \xrays{}
is \num{\sim0.05} times that in the \ac{UV}
\citep[e.g.,][]{1981ApJ...245..357Z}. Indeed, magnetic field of strength
\SI{\sim10}{\milli\gauss} has been detected on \SI{\lesssim30}{\parsec} scales
in the nucleus of \ac{NGC}~1068 \citep{2015MNRAS.452.1902L}. It is therefore
critical that we understand what implications magnetic field has on torus
dynamics.

This article reports the extension of our \acp{RHD} simulations
\citepalias{2016ApJ...825...67C} to \acp{RMHD}. Our code marries the
finite-volume \acp{MHD} code Athena \citep{2008ApJS..178..137S} with
time-dependent \ac{IR} \citep{2014ApJS..213....7J} and time-independent \ac{UV}
\citepalias{2016ApJ...825...67C} \ac{RT} modules. It is uniquely capable of
concurrently solving the \acp{MHD} and \ac{RT} equations without resorting to
arbitrary closures for \ac{RT}; this property is crucial for correctly treating
gas--radiation interaction in systems with optical depths comparable to unity
\citep{2012ApJS..199....9D}.

In addition, we quantitatively explore the effect of varying the rotational
profile of the torus. An important lesson is that a torus with Keplerian
rotation cannot withstand irradiation for a long time (\cref{sec:parameter
study}), but a torus with sub-Keplerian rotation may maintain a steady
inflow--outflow morphology and survive under mass loss for multiple orbits
(\cref{sec:steady state}). The kinematics and obscuration properties of the
radiation-driven outflow in this \textquote{steady} state generally agree with
observations.

We recount our simulation setup in \cref{sec:methods}, report our results in
\cref{sec:UV-RMHD results,sec:IR-RMHD results}, and discuss them in
\cref{sec:discussion}.

\section{Methods}
\label{sec:methods}

Our simulations are based on \citetalias{2016ApJ...825...67C}; the only
differences are in our initial condition (\cref{sec:MHD stage}) and in how we
conduct the simulations (\cref{sec:strategy}). We adopt cylindrical coordinates
$(R,\phi,z)$ and define the spherical radius $r\eqdef(R^2+z^2)^{1/2}$ for
convenience. Quantities are normalized by the fiducial quantities listed in
\cref{tab:fiducial}. The surface on which the \ac{UV} optical depth
$\tau_\su{UV}$ to the central source is unity is called the \textquote{inner
surface}, whereas the part of this surface near the mid-plane is called the
\textquote{inner edge}.

\begin{table}
\caption{Fiducial quantities.}
\label{tab:fiducial}
\begin{tabular}{ccc}
\toprule
Fiducial quantity & Symbol & Definition \\\midrule
temperature & $T_\su{ds}$ & \SI{1500}{\kelvin} \\
opacity per mass & $\kappaT$ & \SI{0.397}{\centi\meter\squared\per\gram} \\
luminosity & $L_\su E$ & $4\pi GMc/\kappaT$ \\
length & $r_0$ & $[L_\su E/(4\pi c\aSB T_\su{ds}^4)]^{1/2}$ \\
velocity & $v_0$ & $(GM/r_0)^{1/2}$ \\
time & $t_0$ & $(GM/r_0^3)^{-1/2}$ \\
gas density & $\rho_0$ & $(\kappaT r_0)^{-1}$ \\
gas pressure & $p_0$ & $\rho_0v_0^2=\aSB T_\su{ds}^4$ \\
magnetic field & $B_0$ & $p_0^{1/2}$ \\
radiation energy density & $E_0$ & $L_\su E/(4\pi r_0^2c)=p_0$ \\
radiative flux & $F_0$ & $cE_0$ \\
\bottomrule
\end{tabular}
\end{table}

\subsection{\texorpdfstring{\Aclp*{MHD}}{Magnetohydrodynamics}}

The equations of ideal \acp{MHD} are
\begin{align}
\label{eq:gas mass}
\pd\rho t+\divg(\rho\vec v) &= 0, \\
\label{eq:gas momentum}
\pd{}t(\rho\vec v)+\divg(\rho\vec v\vec v+p^*\tsr I-\vec B\vec B) &=
  -\rho\grad\Phi+\momsrc{IR}+\momsrc{UV}, \\
\label{eq:gas energy}
\pd Et+\divg[(E+p^*)\vec v-(\vec B\cdot\vec v)\vec B] &=
  -\rho\vec v\cdot\grad\Phi+\ergsrc{IR}+\ergsrc{UV}, \\
\label{eq:induction}
\pd{\vec B}t-\curl(\vec v\cross\vec B) &= \vec0.
\end{align}
Here $\rho$, $\vec v$, and $p$ are gas density, velocity, and pressure. The
magnetic field is $\vec B$, whose unit is chosen such that magnetic
permeability is unity. Gas temperature, total pressure, and total energy
density are $T=p/(\rho R_\su{ideal})$, $p^*=p+\tfrac12B^2$, and $E=\tfrac12\rho
v^2+p/(\gamma-1)+\tfrac12B^2$ respectively, where $R_\su{ideal}$ and $\gamma$
are the specific ideal gas constant and the ratio of specific heats. The
gravitational potential of the central mass is $\Phi(\vec r)=-GM/r$. The energy
and momentum source terms due to radiation are $\ergsrc{IR,UV}$ and
$\momsrc{IR,UV}$, to be defined in \cref{sec:radiative transfer}. Finally, the
isotropic rank-two tensor is denoted by $\tsr I$.

\subsection{\texorpdfstring{\Acl*{RT}}{Radiative transfer}}
\label{sec:radiative transfer}

The torus is illuminated in the \ac{UV} by the innermost regions of an
accretion disk at the origin. The angular distribution of the \ac{UV} radiative
flux is modulated by a number of physical effects: Geometrical projection and
limb darkening due to a scattering atmosphere favor emission in the polar
direction, while relativistic boosting, beaming, and lensing enhance the
equatorial radiative flux. For simplicity, we adopt a spherically symmetric
central source in our simulations.

The propagation of \ac{IR} radiation is handled by the time-dependent \ac{RT}
module of Athena, which solves the time-dependent \ac{RT} equation on a large
number of grid rays. To first order in $v/c$, where $c$ is the speed of light,
the mixed-frame time-dependent \ac{RT} equation for \ac{IR} radiation
interacting with gray material reads \citep{2014ApJS..213....7J}
\begin{multline}\label{eq:IR RT}
\frac1{\hat c}\pd{I_\su{IR}}t+\uvec n\cdot\grad I_\su{IR}=
  \Bigl(-1+\uvec n\cdot\frac{\vec v}c\Bigr)
  \rho (\kappa_\su{IR}+\sigma_\su{IR})I_\su{IR} \\
+\Bigl(1+3\,\uvec n\cdot\frac{\vec v}c\Bigr)
  \rho (\kappa_\su{IR}B+\sigma_\su{IR}J_\su{IR})
  -2\rho\sigma_\su{IR}\frac{\vec v}c\cdot\vec H_\su{IR} \ifaastex\relax\\
+\rho (\kappa_\su{IR}-\sigma_\su{IR})\frac{\vec v}c\cdot
  (\vec H^0_\su{IR}-\vec H_\su{IR}).
\end{multline}
The specific intensity integrated over the \ac{IR} in the observer frame is
$I_\su{IR}(\uvec n)$; its lowest three angular moments are $J_\su{IR}$, $\vec
H_\su{IR}$, and $\tsr K_\su{IR}$, from which the \ac{IR} radiation energy
density and flux follow as $E_\su{IR}=(4\pi/c)J_\su{IR}$ and $\vec
F_\su{IR}=4\pi\vec H_\su{IR}$. The frequency-integrated black-body mean
intensity is $B(T)=c\aSB T^4/(4\pi)$, where $\aSB$ is the radiation constant.
The comoving absorption and scattering cross sections per mass in the \ac{IR}
are $\kappa_\su{IR}$ and $\sigma_\su{IR}$ respectively, and $\hat c$ is the
reduced speed of light
\citetext{\citetalias{2016ApJ...825...67C}\multicitedelim\citealp[see
also][]{2001NewA....6..437G, 2013ApJS..206...21S}}. Taking the zeroth and first
angular moments of \cref{eq:IR RT} yields
\ifaastex{
  \begin{alignat}{2}
  \label{eq:IR RT zeroth moment}
  \frac1{\hat c}\pd{J_\su{IR}}t+\divg\vec H_\su{IR} &=
    \rho\kappa_\su{IR}(B-J_\su{IR})
    +\rho (\kappa_\su{IR}-\sigma_\su{IR})\frac{\vec v}c\cdot\vec H^0_\su{IR}
    &&\eqdef -\frac1{4\pi}\ergsrc{IR}, \\
  \label{eq:IR RT first moment}
  \frac1{\hat c}\pd{\vec H_\su{IR}}t+\divg\tsr K_\su{IR} &=
    \rho\kappa_\su{IR}\frac{\vec v}c(B-J_\su{IR})
    -\rho (\kappa_\su{IR}+\sigma_\su{IR})\vec H^0_\su{IR}
    &&\eqdef -\frac c{4\pi}\momsrc{IR}.
  \end{alignat}
}{
  \begin{alignat}{2}
  \nonumber
  & \frac1{\hat c}\pd{J_\su{IR}}t+\divg\vec H_\su{IR}= \\
  \label{eq:IR RT zeroth moment}
  &\quad \rho\kappa_\su{IR}(B-J_\su{IR})
    +\rho (\kappa_\su{IR}-\sigma_\su{IR})\frac{\vec v}c\cdot\vec H^0_\su{IR}
    &&\eqdef -\frac1{4\pi}\ergsrc{IR}, \\
  \nonumber
  & \frac1{\hat c}\pd{\vec H_\su{IR}}t+\divg\tsr K_\su{IR}= \\
  \label{eq:IR RT first moment}
  &\quad \rho\kappa_\su{IR}\frac{\vec v}c(B-J_\su{IR})
    -\rho (\kappa_\su{IR}+\sigma_\su{IR})\vec H^0_\su{IR}
    &&\eqdef -\frac c{4\pi}\momsrc{IR}.
  \end{alignat}
}
The remaining piece to specify in \cref{eq:IR RT,eq:IR RT zeroth moment,eq:IR
RT first moment} is $\vec H^0_\su{IR}$, the first angular moment of the \ac{IR}
specific intensity in the fluid frame. It is related to the angular moments in
the observer frame by a Lorentz transformation \citep{1984oup..book.....M}:
\begin{equation}
\vec H^0_\su{IR}=\vec H_\su{IR}
  -\frac{\vec v}cJ_\su{IR}-\frac{\vec v}c\cdot\tsr K_\su{IR}+\bigO(v^2/c^2).
\end{equation}

The time-dependent \ac{RT} module of Athena introduces artifacts along grid
rays in optically thin regions; therefore, we have written a time-independent
long-characteristics \ac{RT} module specifically for \ac{UV} radiation from the
central source. The module computes the \ac{UV} radiation energy density in any
cell as
\begin{equation}
\frac{4\pi}cJ_\su{UV}\eqdef\frac{L_\su{UV}}{4\pi r^2c}e^{-\tau_\su{UV}}
  \frac{\exp(\tfrac12\tau^*_\su{UV})-\exp(-\tfrac12\tau^*_\su{UV})}
  {\tau^*_\su{UV}},
\end{equation}
where $L_\su{UV}$ is the luminosity of the central source, while $\tau_\su{UV}$
and $\tau^*_\su{UV}$ are the \ac{UV} optical depths from the source to the cell
and across the cell respectively. The energy and momentum source terms due to
\ac{UV} radiation are then
\begin{align}
\label{eq:UV energy source}
-\frac1{4\pi}\ergsrc{UV} &\eqdef
  -\rho\kappa_\su{UV}J_\su{UV}\Bigl(1-\uvec e_r\cdot\frac{\vec v}c\Bigr), \\
\label{eq:UV momentum source}
-\frac c{4\pi}\momsrc{UV} &\eqdef
  -\rho\kappa_\su{UV}J_\su{UV}\,\uvec e_r\,
  \Bigl(1-\uvec e_r\cdot\frac{\vec v}c\Bigr),
\end{align}
with $\kappa_\su{UV}$ being the comoving absorption cross section per mass in
the \ac{UV}.

The chief sources of opacity in our system are dust absorption and electron
scattering, which we model as
\begin{align}
\kappa_\su{IR}(T) &\eqdef \bar\kappa_\su{IR}\times
  \frac12\left[1-\tanh\frac{\log_{10}(T/T_\su{ds})}{\Delta_\su{ds}}\right], \\
\label{eq:UV opacity}
\kappa_\su{UV}(T) &\eqdef \bar\kappa_\su{UV}\times
  \frac12\left[1-\tanh\frac{\log_{10}(T/T_\su{ds})}{\Delta_\su{ds}}\right], \\
\sigma_\su{IR}(T) &\eqdef \kappaT\times
  \frac12\left[1+\tanh\frac{\log_{10}(T/T_\su{hi})}{\Delta_\su{hi}}\right].
\end{align}
In these fitting formulae, $T_\su{ds}\approx\SI{1500}{\kelvin}$ is the dust
sublimation temperature \citep[e.g.,][]{1969Natur.223..788R,
1981ApJ...250...87R, 1987ApJ...320..537B}, $T_\su{hi}\approx\SI{4013}{\kelvin}$
is the temperature at which hydrogen atoms in \ac{LTE} at a number density of
\SI{e4}{\per\cubic\centi\meter} are collisionally half-ionized, and
$\kappaT\approx\SI{0.397}{\centi\meter\squared\per\gram}$ is the Thomson
scattering cross section per mass. The dust opacities are normalized to Thomson
as $\bar\kappa_\su{IR}/\kappaT=20$ and $\bar\kappa_\su{UV}/\kappaT=80$; the
parameters governing the transition between opacity regimes are
$\Delta_\su{ds}=0.05$ and $\Delta_\su{hi}\approx0.196$.

\subsection{Simulation strategy}
\label{sec:strategy}

\Ac{MHD} turbulence stirred by the \ac{MRI} saturates after several tens of
orbits, but \acp{RMHD} simulations are computationally prohibitively expensive
on that timescale. One solution is to carry out the simulation in stages; in
order of execution, they are the \acp{MHD} stage, the \uvrmhd{} stage, and the
\irrmhd{} stage. Each stage is documented at length below.

\subsubsection{\texorpdfstring{\acsp*{MHD}}{MHD} stage}
\label{sec:MHD stage}

In the \acp{MHD} stage, we follow a geometrically thick, gas-supported torus in
pure \acp{MHD} to saturation. The rationale is that if \ac{IR} radiation does
behave like a pressure in optically thick regions, we should be able to replace
gas pressure by \ac{IR} radiation pressure at a later stage with minimal
changes to the geometry of the torus.

We begin by constructing the initial condition for the \acp{MHD} stage.
Assuming $\vec v(R)=(GM/R_\su p)^{1/2}(R/R_\su p)^{1-q}\,\uvec e_\phi$ and
$p=K\rho^\Gamma$, the density of an axisymmetric hydrostatic torus is uniquely
given by \citep{1984MNRAS.208..721P}
\begin{equation}
\const=\frac{GM}r+\frac{v_\phi^2}{2-2q}
-\begin{cases}
  K\tfrac\Gamma{\Gamma-1}\rho^{\Gamma-1}, & \Gamma\ne1, \\
  K\ln\rho, & \Gamma=1,
\end{cases}
\end{equation}
with maximum $\rho=\rho_\su p$ at $(R,z)=(R_\su p,0)$. The free parameters of
the initial condition are the radial coordinate $R_\su p$ of the density
maximum, the shear parameter $q$, the polytropic constant $K$, the polytropic
index $\Gamma$, and the constant on the left-hand side. Note that we require
$1.5<q<2$ for the torus to have finite height and be stable; indeed, a
geometrically thick torus supported by isotropic pressure must have a
sub-Keplerian rotational profile (\cref{sec:angular momentum}). A geometrically
thick torus with bounded radial and vertical extent can be obtained with $R_\su
p=r_0$, $\rho_\su p=\rho_0$, $q=1.9$, $K=0.2p_0\rho_0^{-\Gamma}$, and
$\Gamma=4$. The surface of the torus is $\rho(R,z)=0$; solving this equation
for $z=0$ yields $R\approx0.545\,r_0$ and $R\approx4.72\,r_0$.

Overlaid on the torus is a poloidal loop of magnetic field derived from the
vector potential $\vec A\propto\max(\rho R^\xi-C\rho_0r_0^\xi,0)\,\uvec
e_\phi$; we choose an exponent $\xi=0.4$ so that the plasma betas at the inner
and outer surfaces of the torus are similar, and a cutoff $C=0.4$ so that all
field lines are properly confined within the torus. The proportionality
constant is selected to make the ratio of volume-integrated gas pressure to
volume-integrated magnetic energy equal to 1000. Gas pressure is also perturbed
at the 0.01 level to seed the \ac{MRI}.

We set up the ambient material around the torus as in
\citetalias{2016ApJ...825...67C}, but here its length and density scales are
$R_\su{amb}=r_0$ and $\bar\rho_\su{amb}=\num{e-5}\,\rho_0$ respectively, and
its sound speed is $v_\phi(R_\su{amb})$.

The configuration is evolved until \ac{MHD} turbulence has fully saturated, at
$t=200\,t_0$. \Ac{MHD} stresses have increased the specific angular momentum
$j\eqdef Rv_\phi$ at $(R,z)=(2\,r_0,0)$ from \num{\approx0.76} to
\num{\approx0.81} times Keplerian, and steepened its mid-plane radial profile
from $j(R)\propto R^{0.1}$ to
\begin{equation}
j(R)\propto\begin{cases}
  R^{0.41}, & R\lesssim2\,r_0, \\
  R^{0.25}, & R\gtrsim2\,r_0.
\end{cases}
\end{equation}

Meanwhile, the torus outer edge spreads radially outward and the torus loses
mass through all boundaries, but eventually the torus stabilizes. To make up
for the mass loss, we multiply $\rho^{1/2}$, $p^{1/2}$, and $\vec B$ at the end
of the \acp{MHD} stage by the same factor everywhere, determined as follows.
The Thomson optical depth averaged over all sightlines contained within a
mid-plane wedge of aspect ratio $m\ll1$ is
\begin{equation}
\mean{\tau_\su T}\approx\frac{(1+m^2)^{1/2}}{2m(\phi_\su{max}-\phi_\su{min})}
  \kappaT\int dV\,\rho r^{-2}\step{mR-\abs z},
\end{equation}
where $\step\cdot$ is the step function and $\phi_\su{\min,\max}$ are the
coordinates of the azimuthal boundaries of the simulation domain. The factor is
chosen such that $\mean{\tau_\su T}$ of a wedge with $m=0.1$ matches some
desired value, which we shall give in \cref{sec:angular momentum}.

\subsubsection{\titleuvrmhd{} stage}
\label{sec:UV-RMHD stage}

We now switch over to \acp{RMHD}. When the central source is turned on and
radiation starts pushing on the gas at the inner surface, the torus is no
longer in equilibrium. Since we would like a steady state, the simulation must
be run long enough that any transients excited at the inner surface have time
to propagate radially outward, away from the inner surface where the dynamics
is the most interesting. The \uvrmhd{} stage accomplishes this: We run the
\acp{MHD} solver with the \ac{UV} \ac{RT} module but not the \ac{IR} \ac{RT}
module. The most expensive step of the \ac{UV} \ac{RT} module is ray casting,
which is performed just once before the simulation starts, thus the amortized
cost is low. In addition to being a preprocessing step for the \irrmhd{} stage,
the \uvrmhd{} stage is also valuable for studying dynamics driven exclusively
by \ac{UV} radiation.

The \ac{UV} \ac{RT} module needs to be modified specifically for this stage.
First, gas temperature in this stage is unsuitable for computing
$\kappa_\su{UV}$ with \cref{eq:UV opacity} because gas temperature in a
gas-supported torus is virial, whereas gas temperature in a realistic
\ac{IR}-supported torus reflects the balance between radiative absorption and
re-emission. Since \ac{UV} radiative acceleration is strong only in the central
hole, we estimate what the gas temperature there may be in the \irrmhd{} stage,
use that expressly for $\kappa_\su{UV}$ in the \uvrmhd{} stage, while keeping
the actual gas temperature unchanged. As we shall justify in
\cref{sec:temperature}, such an opacity temperature would be
\begin{equation}\label{eq:opacity temperature}
T_\su{op}(r)\eqdef\left(\frac{\bar\kappa_\su{UV}L_\su{UV}}
  {4\pi\bar\kappa_\su{IR}c\aSB r^2}\right)^{1/4};
\end{equation}
this means $\kappa_\su{UV}$ is a function not of gas temperature, but of
position. Second, gas should convert most of the energy it receives from
\ac{UV} radiation to the \ac{IR}, but since it cannot do so without the \ac{IR}
\ac{RT} module, \cref{eq:UV energy source} overestimates the energy actually
imparted to the gas. The correct value of $\ergsrc{UV}$ should simply be the
rate of work done by \ac{UV} radiation, that is, $\ergsrc{UV}\eqdef\vec
v\cdot\momsrc{UV}$. Note that although \citet{2012ApJ...758...66W} also
considered the case where \ac{UV} radiation from the central source deposits
momentum, not energy, their energy equation does not include a similar term.
Third, we assume gas velocity vanishes in \cref{eq:UV momentum source}, thereby
ignoring the minute effect of Lorentz transformation. Fourth, the time step is
arbitrarily multiplied by 0.25 to account for the fact that $\momsrcmag{UV}\gg
GM\rho/r^2$ at $\tau_\su{UV}\lesssim1$.

We reset $t$ to zero when the \uvrmhd{} stage begins; consequently, all times
reported below are reckoned from the beginning of the stage.

\subsubsection{Reduction of angular momentum in \titleuvrmhd{} stage}
\label{sec:angular momentum}

\Ac{MHD} stresses establish $j(R)$ in realistic tori. Simulating this process
all the way to steady state is impossible in the \irrmhd{} stage because of
computational cost; it is also impossible in either the \uvrmhd{} stage or the
\irrmhd{} stage because \ac{MHD} stresses redistribute angular momentum over
tens of orbits, but our choice of $\kappa_\su{UV}$ implies that the
radiation-driven outflow drains all the mass from an isolated torus in a couple
orbits, and our simulations do not provide continuous mass resupply. Although
the torus cannot reach a formal steady state in our simulations, it may
nonetheless exhibit an approximate \textquote{steady} state wherein its inner
edge stays close to the dust sublimation surface and its morphology is
qualitatively the same over time. This \textquote{steady} state ends if \ac{UV}
radiation does enough positive work over time to gravitationally unbind the
torus, but the amount of work required depends on $j(R)$. We may obtain a
relatively long-lasting \textquote{steady} state by reducing $j(R)$ of the
\ac{MHD} stage output before forwarding it to the \uvrmhd{} stage, but we must
first determine what $j(R)$ produces the longest \textquote{steady} state.

\Citet{2011ApJ...741...29D} likewise concluded that geometrically thick tori
must have sub-Keplerian rotation, but on the basis of maintaining dynamical
equilibrium in the spherically radial direction. Their argument can be recast
more generally: Isotropic \ac{IR} radiation pressure provides vertical and
radial support simultaneously, hence a geometrically thick, \ac{IR}-supported
torus can only be in radial balance if rotation is sub-Keplerian. The same
logic applies whenever geometrical thickness is ascribed to some isotropic
pressure, be it gas pressure, radiation pressure, magnetic pressure, or
velocity dispersion. Similar to ours, the latest simulations by
\citet{2016ApJ...819..115D} employed an initial condition whose angular
momentum profile is shallower than Keplerian.

Irradiation strengthens this argument. A geometrically thick torus, by
definition, intercepts a sizable fraction of the radiation from the central
source. Some of the radiative momentum absorbed by the torus may be carried
away in an outflow, but whatever left behind constitutes a radially outward
force, which could be strong enough to counteract gravity if $L/L_\su
E\gtrsim\tau_\su T$. Consider an unirradiated torus in which rotation precisely
balances gravity; in other words, rotation is sub-Keplerian only to the extent
isotropic pressures, if present, compel it to be. This torus cannot remain in
equilibrium when the central source is turned on; to do so, its rotation must
be even more sub-Keplerian. Moreover, radiation does positive work on
outward-moving gas; if this energy is not advected away in its entirety by the
outflow, the torus will increase in mechanical energy and be unbound
eventually. The torus can stay in place only if it is replenished with mass to
compensate for the outflow, and if this mass has sub-Keplerian rotation to
offset the gain in mechanical energy.

For ease of parameterization, we multiply $j(R)$ from the \acp{MHD} stage by
$\alpha (R/r_\su{ds})^{-\beta}$, where $0<\alpha\le1$, $\beta\ge0$, and the
dust sublimation radius is defined by \citepalias{2016ApJ...825...67C}
\begin{equation}\label{eq:dust sublimation radius}
r_\su{ds}^2\eqdef
  \frac{\kappa_\su{UV}L_\su{UV}}{4\pi\kappa_\su{IR}c\aSB T_\su{ds}^4}.
\end{equation}
A parameter study decides the optimal values of $\alpha$ and $\beta$. While the
computational cost of the \uvrmhd{} stage is merely a few percent of the
\irrmhd{} stage, extensive sampling of the parameter space is still unfeasibly
expensive; we have therefore tested 11 pairs of parameters, as shown in the
left panel of \cref{fig:UV-RMHD parameter study}. All runs have the same
gravitational potential energy $E_\su{grav}$ at $t=0$ but different kinetic
energy $E_\su{kin}$; consequently, a useful parameter is
$b\eqdef1+E_\su{kin}/E_\su{grav}$, the binding energy normalized to the
negative of the gravitational potential energy. Conscious effort is expended to
ensure each run has a value of $b$ similar to that of at least one other run.
We choose $\mean{\tau_\su T}=1$ for the parameter study because this optical
depth lies within the observed range \citep[e.g.,][]{1999ApJ...522..157R}, but
its exact value is immaterial as long as the torus is optically thick to
\ac{UV} radiation. We shall demonstrate in \cref{sec:parameter study} that
$\alpha=0.8$ and $\beta=0.25$ grant the torus the longest \textquote{steady}
state.

\begin{figure*}
\includegraphics{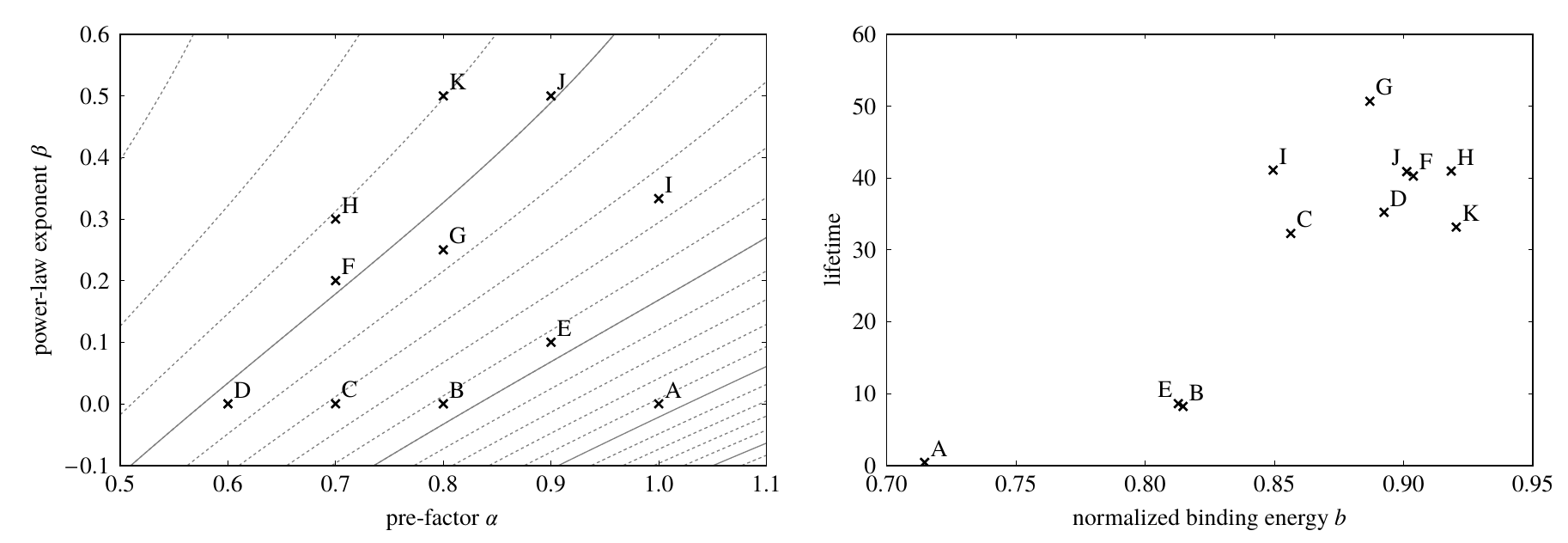}
\caption{\textit{Left panel:} Plot of pre-factor $\alpha$ and power-law
exponent $\beta$ that specify how $j(R)$ is modified just before the \uvrmhd{}
stage. Contours plot $b$, with solid contours at 0.6 to 0.9 from bottom-right
to top-left in steps of 0.1. \textit{Right panel:} Plot of lifetime, normalized
to fiducial units (\cref{tab:fiducial}), against $b$. See \cref{sec:angular
momentum} for the definitions of $j(R)$, $\alpha$, $\beta$, and $b$, and
\cref{sec:parameter study} for the definition of lifetime.}
\label{fig:UV-RMHD parameter study}
\end{figure*}

With parameters thus fixed, it remains to choose a snapshot of the \uvrmhd{}
stage for passing on to the \irrmhd{} stage. We impose two criteria on such
snapshot. First, in the \uvrmhd{} stage, gas falls radially inward due to
decreased rotational support, then rebounds and relaxes upon encountering the
centrifugal barrier. The infalling and relaxing regions are separated by an
outward-propagating shock; we consider a snapshot eligible only if the shock
has moved sufficiently far away from the inner surface. Second, to ensure that
the torus would not be blown away immediately in the \irrmhd{} stage, we
stipulate that the torus survive in the \uvrmhd{} stage beyond the selected
snapshot for two more orbits at the inner edge. A more massive torus is able to
withstand \ac{UV} irradiation longer, allowing ourselves greater freedom in
picking a snapshot that obeys both requirements; therefore, we conduct yet
another run in the \uvrmhd{} stage with the optimal $\alpha$ and $\beta$, but
with $\mean{\tau_\su T}=2$. The snapshot we opt for in the additional run is
$t=50\,t_0$.

This exercise provides us with a \textquote{steady}-state torus in the
\uvrmhd{} stage; of course, there is no guarantee that it would remain so in
the following \irrmhd{} stage.

\subsubsection{\titleirrmhd{} stage}
\label{sec:IR-RMHD stage}

As we advance to the \irrmhd{} stage, we reinstate \ac{IR} radiative support by
replacing gas pressure with a combination of gas and \ac{IR} radiation pressure
under thermal equilibrium; in other words, if gas temperatures before and after
the replacement are $T_1$ and $T_2$ respectively, and the isotropic \ac{IR}
specific intensity in the fluid frame after the replacement is $I^0_2$, then we
demand $\rho R_\su{ideal}T_1=\rho R_\su{ideal}T_2+\tfrac13\aSB T_2^4$ and $\aSB
T_2^4=(4\pi/c)I^0_2$. The degree to which gas pressure is replaced is
quantified by $T_2/T_1=[1+\tfrac13\aSB T_2^4/(\rho R_\su{ideal}T_2)]^{-1}$. We
clearly have $R_\su{ideal}T_2\sim c_\su s^2$, where $c_\su s$ is the sound
speed after the replacement. Since $\aSB T_2^4\sim\rho v_\phi^2$, and since
$c_\su s^2\ll v_\phi^2$ if the torus is to be \ac{IR}-supported and not
gas-supported, we have $T_2/T_1\sim c_\su s^2/v_\phi^2\ll1$.

Several comments are in order. First, the replacement preserves pressure, not
energy or momentum, because we are interested in how the torus is supported.
Second, the replacement does not promise exact force balance in the inertial
frame; in fact, considering that the gas pressure tensor in the fluid frame is
isotropic while the \ac{IR} radiation pressure tensor in the same frame is
ellipsoidal, there is no trivial transformation from one kind of pressure to
another that would secure force balance everywhere. Third, the assumption that
\ac{IR} radiation pressure acts like a gas pressure is valid only in optically
thick regions, so we are not justified to perform the replacement within the
central hole; nevertheless, since the gas there is optically thin to the
central source, the steady-state temperature profile quickly establishes itself
no matter what the initial temperature is.

At the beginning of the stage, the radial Thomson and \ac{IR} optical depths
along the mid-plane are \num{\approx1.59} and \num{\approx23.1}; the mass in
the quarter-circle simulation domain (\cref{sec:parameters}) is
$\mathrelp\approx8.83\times\frac12\pi\rho_0r_0^3$, compared to
$\mathrelp\approx2.10\times\frac12\pi\rho_0r_0^3$ for the initial condition of
the \ac{RHD} torus \citepalias{2016ApJ...825...67C}. We again reset $t$ to
zero. Now that the torus is \ac{IR}-supported, we enable both \ac{IR} and
\ac{UV} \ac{RT} modules and study whether it can self-consistently stay so. The
simulation is conducted in this stage to $t=14\,t_0$. The final radial Thomson
and \ac{IR} optical depths along the mid-plane are $\mathrelp\approx2.34$ and
$\mathrelp\approx37.9$; radiation-driven mass loss results in a final mass of
$\mathrelp\approx6.62\times\frac12\pi\rho_0r_0^3$.

\subsection{Simulation parameters and domain}
\label{sec:parameters}

The simulation domain spans
$[0.3\,r_0,9.9\,r_0]\times[-\tfrac14\pi,\frac14\pi]\times[-5\,r_0,5\,r_0]$ in
$(R,\phi,z)$ in all stages. A large radial extent is needed to capture the
extended flow after the radial expansion of the torus in the \acp{MHD} stage
(\cref{sec:MHD stage}); the geometrical thickness of the torus demands a
similarly large vertical extent. We pick the number of grid cells to be
$480\times60\times500$ in $(R,\phi,z)$, large enough to resolve both
\ac{MRI}-driven turbulence \citep[e.g.,][]{2013ApJ...772..102H} and the \ac{UV}
absorption layer at the inner surface. The number of grid rays per cell is 168.

To make contact with our previous simulations \citepalias{2016ApJ...825...67C},
we let $L_\su{UV}/L_\su E=0.1$ in both \uvrmhd{} and \irrmhd{} stages.
Numerical artifacts can appear if rotation does not exactly cancel gravity;
such artifacts are smoothed out when $c_\su s/v_\phi\gtrsim\bigO(0.1)$, where
$c_\su s$ and $v_\phi$ are the gas sound speed and orbital speed respectively
at the inner edge \citepalias{2016ApJ...825...67C}. We set up our simulations
such that the gas--radiation equilibrium temperature at the inner edge is
always $\mathrelp\approx T_\su{ds}$, hence $c_\su s$ is independent of $M$.
Since $v_\phi^2\propto M/r_\su{ds}\propto M/L_\su{UV}^{1/2}$, where we used
$r_\su{ds}$ from \cref{eq:dust sublimation radius}, we have
\begin{equation}\label{eq:mass and sound speed}
M\propto(c_\su s/v_\phi)^{-4};
\end{equation}
this means a constraint on $c_\su s/v_\phi$ is also a constraint on $M$. Here
we choose $c_\su s/v_\phi=0.1$ as in \citetalias{2016ApJ...825...67C},
corresponding to $M\approx\SI{0.758}{\solarmass}$. Its actual value in the
\irrmhd{} stage ranges from $c_\su s/v_\phi\approx0.15$ at
$(R,z)=(r_\su{ds},0)$ to $c_\su s/v_\phi\approx0.5$ at $(R,z)=(3.5\,r_0,0)$
(see also \cref{sec:forces}); therefore, gas pressure remains a minor
contributor to support where gas is most dense (\cref{sec:steady state}), and
our results are not qualitatively affected by the small value of $M$.
Quantitative changes may occur at larger $M$ (\cref{sec:extrapolation}), but
\cref{eq:mass and sound speed} shows that the ratio of simulated to realistic
value of $M$ is much smaller than the analogous ratio for $(c_\su
s/v_\phi)^{-1}$.

\subsection{Scaling properties}

Let us examine the scaling properties of \cref{eq:gas mass,eq:gas
momentum,eq:gas energy,eq:induction} in the three stages. In the \acp{MHD}
stage, the radiative source terms are zero; if we adopt a system of
normalization in which $v_0^2=GM/r_0$, it is clear that the dimensionless
equations are independent of $M$. The \uvrmhd{} stage introduces the \ac{UV}
radiative source terms in their modified forms (\cref{sec:UV-RMHD stage}).
Since the normalization of $\momsrc{UV}$ is $\rho_0\kappaT E_0=\rho_0v_0^2/r_0$
(\cref{tab:fiducial}), and the normalization of $\ergsrc{UV}$ is $v_0$ times
that, the dimensionless equations remain independent of $M$. This means we are
not committed to a particular value of $M$ in either stage, and we may simply
scale our results as needed to match any $M$.

The situation is very different in the \irrmhd{} stage. The normalization of
$\momsrc{IR,UV}$ is $\rho_0v_0^2/r_0$ as before, but now $\momsrc{IR,UV}$ has
additional terms beyond zeroth order in $v_0/c$; worse still, the normalization
of $\ergsrc{IR,UV}$ is now $c/v_0$ times that of $\momsrc{IR,UV}$. The
introduction of a fixed velocity scale $c$ therefore breaks scalability in all
equations except \cref{eq:gas mass} and the leading order of \cref{eq:gas
momentum}. Because $v_0^4=GM\kappaT\aSB T_\su{ds}^4$, a choice of $v_0/c$ is
equivalent to a choice of $M$, which we made in \cref{sec:parameters}.

\section{Results of UV-RMHD stage}
\label{sec:UV-RMHD results}

\subsection{\texorpdfstring{\acs*{UV}}{UV}-driven dynamics}
\label{sec:UV dynamics}

All runs except run~A involve an initial suppression of angular momentum
(\cref{sec:angular momentum}) and evolve in qualitatively similar fashion. We
illustrate this general behavior with run~G, displayed in the top row of
\cref{fig:overview}.

\begin{figure*}
\includegraphics{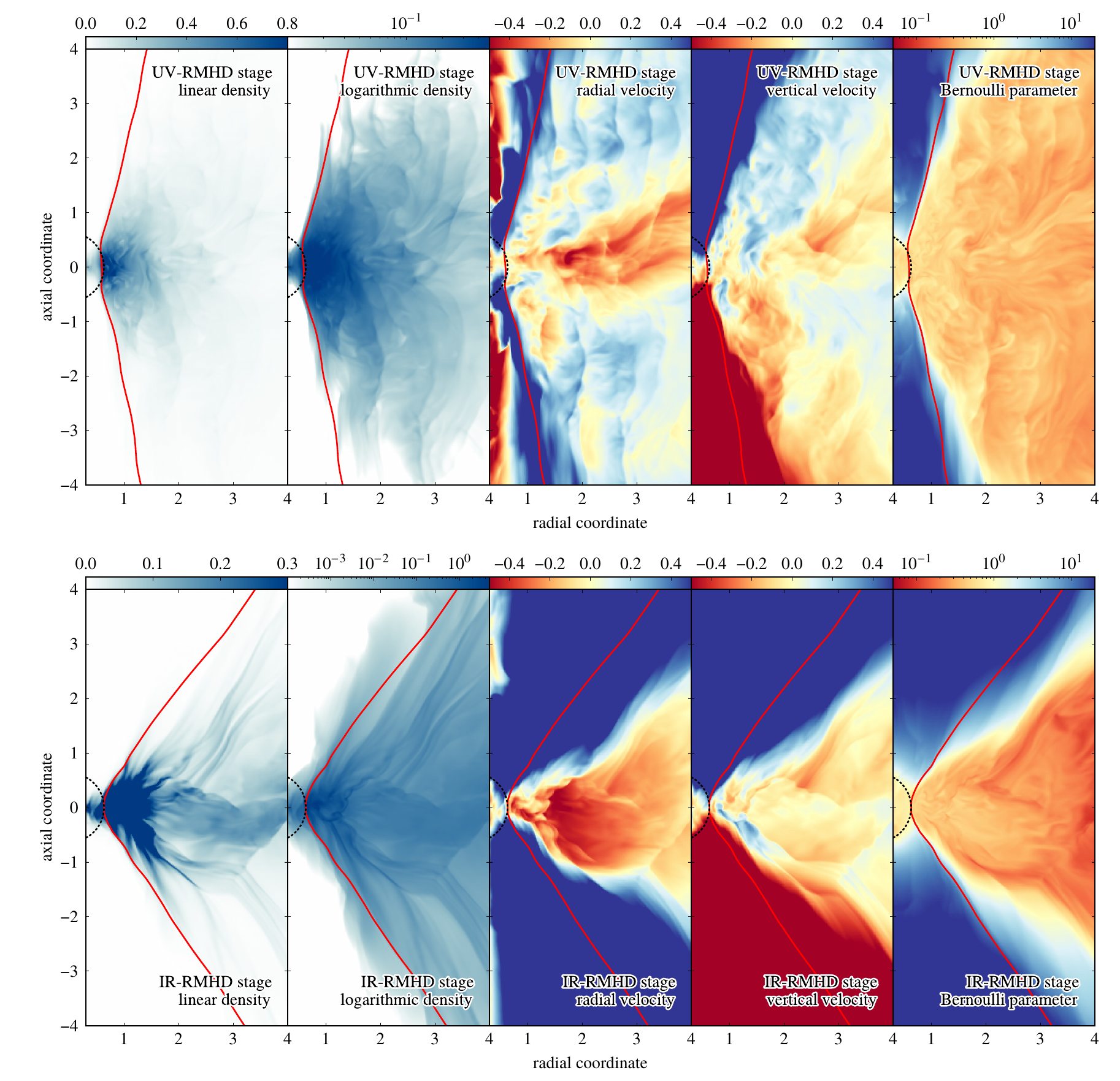}
\caption{Zoom-in of the poloidal plane along $\phi=0$. The dust sublimation
surface $r=r_\su{ds}$ (\cref{eq:dust sublimation radius}) is the dotted black
contour around the origin, and the red contour traces the surface on which
$\tau_\su{UV}=1$. All quantities are normalized to fiducial units
(\cref{tab:fiducial}). \textit{Top row:} Run G in the \uvrmhd{} stage at
$t=40\,t_0$. \textit{Bottom row:} \irrmhd{} stage at $t=14\,t_0$. \textit{First
and second columns:} Gas density is presented on linear and logarithmic scales
respectively as blue intensities (see color bars along the top edge).
\textit{Third to fifth columns:} Colors represent $v_R$, $v_z$, and
$(GM\rho/r)^{-1}(\tfrac12\rho v^2+p)$ respectively (see color bars along the
top edge).}
\label{fig:overview}
\end{figure*}

The torus quickly settles into a \textquote{steady} state (\cref{sec:angular
momentum}). Because rotation alone provides insufficient support against
gravity, gas falls radially inward at speeds $\mathrelp\sim v_\phi$ and
converges toward the inner edge. After passing through a shock, the gas joins a
\definition{lump} at $0.6\,r_0\lesssim R\lesssim r_0$ and $\abs
z\lesssim0.3\,r_0$, highlighted in \cref{fig:cartoon}. \Ac{UV} radiation opens
up the central hole as expected, but only weakly; all runs considered, the
\ac{UV} half--opening angle, defined as the angle between the axis and the
$\tau_\su{UV}=1$ surface, finds equilibrium somewhere between
\SI{\sim0.24}{\radian} and \SI{\sim0.38}{\radian}.

\begin{figure}
\includegraphics{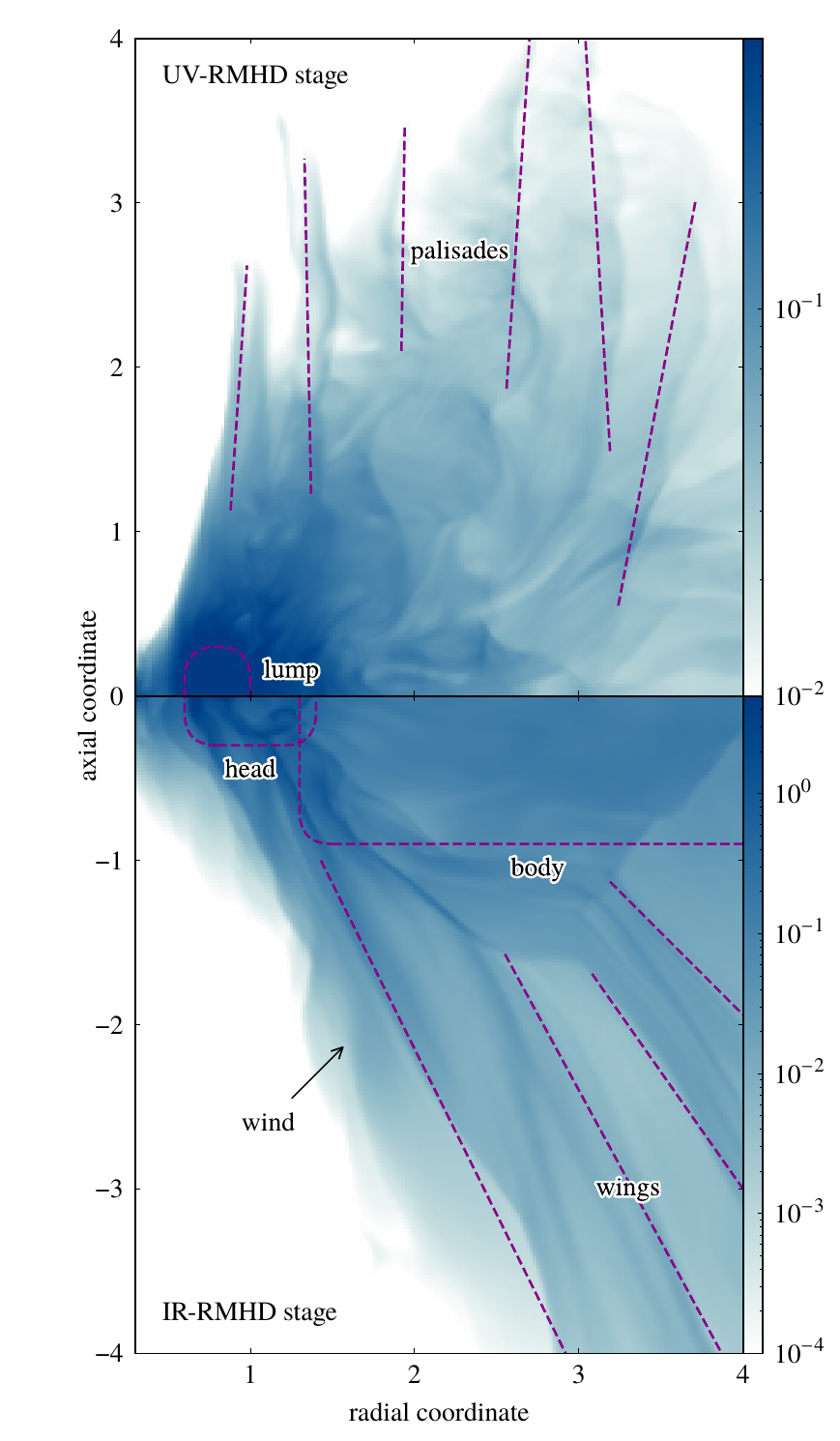}
\caption{Persistent structures sketched on top of the gas density plot taken
from the second column of \cref{fig:overview}. The structures are discussed in
\cref{sec:UV dynamics,sec:steady state}.}
\label{fig:cartoon}
\end{figure}

Wind launching by \ac{UV} radiation is bursty. When gas is shot out from the
inner edge, it is accelerated along the $\tau_\su{UV}=1$ surface by \ac{UV}
radiation, and at the same time pushed horizontally outward beyond the
$\tau_\su{UV}=1$ surface by centrifugal and radiative accelerations. Such gas
excites a weak shock that propagates outward into the slower-moving gas at
$\tau_\su{UV}\gtrsim1$, a shock we call a \definition{palisade}.
\Cref{fig:cartoon} has palisades at $\abs z/R\gtrsim0.5$ and
$\tau_\su{UV}\gtrsim1$; \cref{fig:palisade} shows a schematic palisade.
Palisades are found exclusively above a certain height because the lump stops
outward motion at low latitudes.

\begin{figure}
\includegraphics{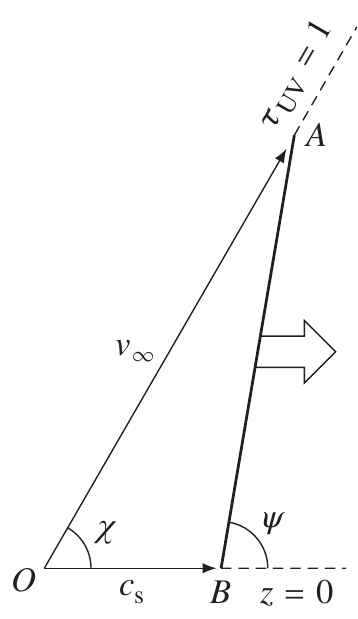}
\caption{Schematic diagram of a palisade in poloidal section. Gas is launched
from the inner edge $O$ along the inner surface $OA$ at characteristic speed
$v_\infty$; as it travels, it excites a shock $AB$ that propagates outward into
the torus at speed $c_\su s$. Palisades appear vertical because
$\psi\approx\tfrac12\pi\,\si{\radian}$ for our parameters.}
\label{fig:palisade}
\end{figure}

The angle $\psi$ between the palisades and the mid-plane is fixed by three
parameters: the angle the $\tau_\su{UV}=1$ surface makes with the mid-plane,
$\chi\sim\SI{1.3}{\radian}$; the characteristic speed of \ac{UV}-launched gas,
$v_\infty\sim(GM/r_\su{ds})^{1/2}(L_\su{UV}/L_\su
E)^{1/2}(\bar\kappa_\su{UV}/\kappaT)^{1/2}$ \citepalias{2016ApJ...825...67C};
and the shock propagation speed in the gas-supported torus, which is the sound
speed $c_\su s\sim(GM/r_\su{ds})^{1/2}$. The angles in \cref{fig:palisade} are
related by \begin{equation}\label{eq:wing angle}
\frac{v_\infty}{\sin(\pi-\psi)}\approx\frac{c_\su s}{\sin(\psi-\chi)};
\end{equation} for our parameters (\cref{sec:radiative
transfer,sec:parameters}), we have $c_\su s/v_\infty\approx0.354$, so the
solution to the equation is $\psi\approx\SI{1.6}{\radian}$. Palisades therefore
look vertical in our simulations, but they need not be so in simulations with
different parameters.

Palisades are important because they represent a sizable portion of the
outflow: They have $v_r>0$ even at $\tau_\su{UV}\gg1$, and most of the mass
outflow through the vertical boundaries in the \textquote{steady} state is in
fact at $\tau_\su{UV}\gtrsim1$. Furthermore, their immediate adjacency to
infalling gas implies inflow and outflow could interact, or even regulate each
other.

\Cref{fig:overview} also shows $(GM\rho/r)^{-1}(\tfrac12\rho v^2+p)$, which is
related to the Bernoulli constant; roughly speaking, gas with this parameter
below or above unity is gravitationally bound to or unbound from the system
respectively. The palisades are bound, and \ac{UV} radiation cannot unbind them
since they are at $\tau_\su{UV}\gtrsim1$, so the outflow in the palisades
cannot continue indefinitely outward.

\subsection{Parameter study of reduction of angular momentum}
\label{sec:parameter study}

One requirement for \textquote{steady} state is that the inner edge stays near
the dust sublimation surface (\cref{sec:angular momentum}). We therefore define
\definition{lifetime} as the time when the intersection of the azimuthally
averaged $\tau_\su{UV}=1$ surface and the mid-plane passes outside
$R=r_\su{ds}$ for the last time; we use the last time because the position of
the inner edge fluctuates at the beginning and its initial crossings of the
dust sublimation surface are of little import. It is unsurprising that lifetime
increases with $b$, as evidenced by the right panel of \cref{fig:UV-RMHD
parameter study}. Variation in $\alpha$ or $\beta$ mostly creates scatter
around this trend.

All runs have the same $L_\su{UV}$, while the \ac{UV} covering fraction
$C_\su{UV}$, defined as the solid angle around the central source with \ac{UV}
optical depth greater than unity, changes by at most a few percent over time
and from run to run; therefore, the rate of \ac{UV} momentum deposition also
varies by a similar amount. We might expect lifetime to increase linearly with
$b$, but that thought is not corroborated by the right panel of
\cref{fig:UV-RMHD parameter study}. A plausible explanation is that the rate of
work done by \ac{UV} radiation is $\vec v\cdot(\rho\kappa_\su{UV}\vec
F_\su{UV}/c)$, so halting infall in fact raises binding energy at a rate
proportional to $-v_r$. A torus with lower $j(R)$ has decreased radial support
and faster inflow, hence \ac{UV} radiation is less effective at unbinding it.

Lifetime diminishes for $b\gtrsim0.9$ because such large binding energy is due
to a $j(R)$ so far below Keplerian that \ac{UV} radiation cannot prevent a
large portion of the gas from falling all the way through the inner-radial
boundary of the simulation domain. Consequently, runs~H and K cannot represent
obscuring tori.

Of the remaining runs, run~G stands out with the longest lifetime; we therefore
select its parameters for the additional run that eventually goes on to the
\irrmhd{} stage (\cref{sec:angular momentum}). Incidentally, run~G at $t=0$ has
approximately flat mid-plane $j(R)$. \Cref{fig:angular momentum} shows that
$j(R)$ stays flat in the mean throughout the \uvrmhd{} stage, yet its
increasing jaggedness suggests that the torus is moving away from a steady
state, namely, that of the \acp{MHD} stage. Because specific angular momentum
is nearly homogeneous at $t=0$, this must be due to angular momentum
redistribution by either \ac{MHD} stresses or non-axisymmetric pressure
fluctuations. It is probable, but by no means certain, that the \uvrmhd{} stage
has its own steady state, and the torus must pass through a disturbed state to
reach it, but our simulations are not long enough for this to happen.

\begin{figure}
\includegraphics{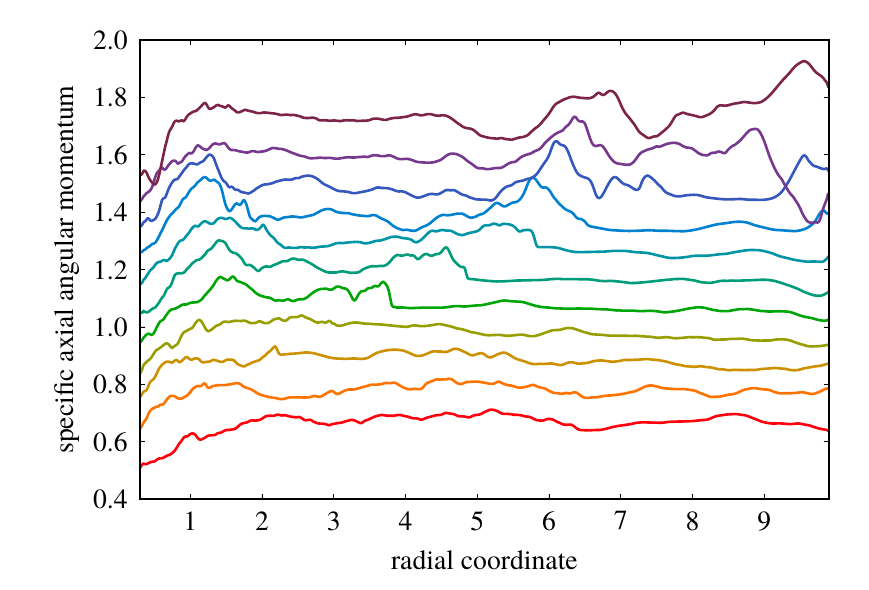}
\caption{Plot of specific axial angular momentum, weighted by density and
averaged azimuthally and vertically over $\abs z\le0.5\,r_0$, against radial
coordinate in run~G in the \uvrmhd{} stage. The lines plot snapshots $5\,t_0$
apart, going from $t=0$ at the bottom to $t=50\,t_0$ at the top. Upward shifts
of $0.1\,r_0v_0$ per line are added for legibility. All quantities are
normalized to fiducial units (\cref{tab:fiducial}).}
\label{fig:angular momentum}
\end{figure}

Our parameter study qualitatively corroborates our claim in \cref{sec:angular
momentum} that a steady-state irradiated torus rotates more slowly than one
unirradiated. The exact amount by which rotation is slower can depend on many
factors, likely including $L_\su{UV}/L_\su E$, which we vary and which sets
both the mass loss and the energy gain rates.

\section{Results of IR-RMHD stage}
\label{sec:IR-RMHD results}

Our principal results come from the \irrmhd{} stage. The first part of this
section deals with the initial transient phase (\cref{sec:transient}), but the
rest is dedicated to the \textquote{steady} state: the internal structure of
the torus (\cref{sec:steady state}); the relative importance of different
forces (\cref{sec:forces}); the flow resulting from these forces
(\cref{sec:streamlines}); the character of the radiation-driven, high-latitude,
wide-angle outflow (\cref{sec:outflow}); the distribution of \ac{IR} radiation
(\cref{sec:IR}); the distribution of temperature (\cref{sec:temperature}); and
the distribution of magnetic field (\cref{sec:magnetic field}).

\subsection{Transient behavior}
\label{sec:transient}

The central hole opens up quite dramatically at the beginning of the \irrmhd{}
stage, in contrast to the \uvrmhd{} stage (\cref{sec:UV dynamics}). The
difference can be explained by considering radial force balance in the plane,
say, $z\sim3\,r_0$, near where it intersects the $\tau_\su{UV}=1$ surface. In
the \uvrmhd{} stage, the locus of intersection fluctuates over time but is
generally within $0.5\,r_0\lesssim R\lesssim r_0$. An approximate balance
obtains in that stage at $R\lesssim2\,r_0$ between radially outward centrifugal
and \ac{UV} radiative accelerations, inward gravitational acceleration, and
largely inward gas pressure acceleration. Gravitational and \ac{UV} radiative
accelerations play minor roles here because their radial components, which are
$\mathrelp\propto R/(R^2+z^2)^{3/2}$, peak at $\abs z/R=\smash{\sqrt2}$ for any
given $\abs z$, but $\abs z/R\gg\smash{\sqrt2}$ for the situation under
discussion. Just before the \irrmhd{} stage, most of the gas pressure is
swapped for \ac{IR} radiation pressure. An inward acceleration is accordingly
removed, but it is not replaced because \ac{IR} radiative flux diffuses outward
from $\tau_\su{UV}\lesssim1$; centrifugal acceleration is no longer opposed, so
the central hole widens. As the \ac{UV} half--opening angle increases, the
radial component of \ac{UV} radiative acceleration at $\tau_\su{UV}\sim1$
becomes stronger, hence \ac{UV} radiation begins to participate in expanding
the central hole as well.

Analogous to previous simulations \citepalias{2016ApJ...825...67C}, we observe
a strong chevron-shaped transient propagating radially outward from the inner
surface as the central hole opens up. The transient can be regarded as dividing
the torus into two parts: The gas outside has not fully responded to the change
induced by turning on the \ac{IR} \ac{RT} module, whereas the gas inside has
undergone at least partial relaxation and is evolving toward a
\textquote{steady} state.

\subsection{\texorpdfstring{\textquote{Steady}}{“Steady”}-state behavior}
\label{sec:steady state}

The torus achieves a \textquote{steady} state at $t\gtrsim6\,t_0$ that lasts
until the end of the simulation; at the average mass loss rate during the
\textquote{steady} state, it would survive for \num{\approx21} orbits. The much
longer lifetime compared to \citetalias{2016ApJ...825...67C} is primarily
because the current torus has \num{\gtrsim4} times the mass.

The bottom row of \cref{fig:overview} portrays the torus at a time when the
transient has left the plotted volume entirely. Four structures with
distinctive and persistent morphologies emerge; \cref{fig:cartoon} points out
where they are. The structures are most recognizable in the first and second
panels of \cref{fig:overview}, but the other panels provide kinematic
information that helps demarcate them. One is the \ac{UV}-launched
\definition{wind} at $\tau_\su{UV}\lesssim1$ \citepalias{2016ApJ...825...67C}.
Because the second panel resembles a top-down view of a bird flying toward the
left, we name the other three by analogy with avian anatomy: The
\definition{head} refers to the very dense region enclosed by $0.6\,r_0\lesssim
R\lesssim1.4\,r_0$ and $\abs z\lesssim0.3\,r_0$, the \definition{body} is the
somewhat less dense region at $1.3\,r_0\lesssim R\lesssim4.2\,r_0$ and
$-0.6\,r_0\lesssim z\lesssim0.4\,r_0$, and the \definition{wings} are the
density ridges parallel to the $\tau_\su{UV}=1$ surface at
$\tau_\su{UV}\gtrsim1$ and $\abs z\gtrsim0.5\,r_0$. We shall see in
\cref{sec:streamlines} that these structures are not hydrostatic, but merely
parts of a global flow that retain their shapes throughout the simulation. The
body does not lie entirely along the mid-plane because \ac{MHD} turbulence
breaks the symmetry about it. The head is denser than the lump in the \uvrmhd{}
stage (\cref{sec:UV dynamics}) by a factor of \num{\sim2}; although the head
and the body are the densest parts of the torus, they take up
\SI{\lesssim30}{\percent} of the total mass owing to their small volumes.

Although the torus is already in \textquote{steady} state, the wind is launched
in bursts because density is not smoothly distributed at the inner edge of the
head. This irregularity gives rise to the complex density structure at
$\tau_\su{UV}\lesssim1$. Because the position of the $\tau_\su{UV}=1$ surface
is easily influenced by the presence of trace amounts of dusty gas in the
central hole, the \textquote{steady}-state \ac{UV} half--opening angle
fluctuates between \SI{\approx0.61}{\radian} and \SI{\approx0.69}{\radian} at
$t\gtrsim6\,t_0$ (see also \cref{sec:obscuration}). The average angle,
\SI{\approx0.65}{\radian}, is tantalizingly close to
$\arctan(1/\smash{\sqrt2})\approx\SI{0.62}{\radian}$. We speculate this is
because the cylindrically radial component of \ac{UV} radiative acceleration
attains its maximum in any horizontal slice at $\abs z/R=\smash{\sqrt2}$.

The \textquote{steady}-state inner surface is corrugated in the azimuthal
direction, similar to our simulations in \citetalias{2016ApJ...825...67C}.
Although the corrugation in \citetalias{2016ApJ...825...67C} increased rapidly
in radial span, the one here remains bounded within $0.62\,r_0\lesssim
R\lesssim0.75\,r_0$. We speculate that the corrugation grows only after the
inner edge has started moving away from the dust sublimation surface. This
speculation is supported by the rapid growth of the corrugation in every run of
the \uvrmhd{} stage, but only during this outward recession phase, whereas the
inner edge in the \irrmhd{} stage never separates from the dust sublimation
surface. We further hypothesize that if a realistic torus could maintain a true
steady state as a result of mass resupply (\cref{sec:angular momentum}), its
inner edge may stay close to the dust sublimation surface and hence not have
strong corrugation.

\subsection{Forces}
\label{sec:forces}

The central unanswered question about obscuring tori is the nature of the
forces supporting them vertically against gravity. Our simulation allows all
relevant forces to be measured. We quantitatively define the advective,
centrifugal, gravitational, gas, magnetic, \ac{IR}, and \ac{UV} forces as
\begin{align}
\vec f_\su{adv} &\eqdef -\rho (\vec v\cdot\grad)\vec v, \\
\label{eq:centrifugal force}
\vec f_\su{cent} &\eqdef \frac{\rho v_\phi^2}R\,\uvec e_R, \\
\vec f_\su{grav} &\eqdef -\frac{GM\rho}{r^2}\,\uvec e_r, \\
\label{eq:pressure force}
\vec f_\su{gas} &\eqdef -\grad p, \\
\vec f_\su{mag} &\eqdef -\vec B\cross(\curl\vec B), \\
\label{eq:IR force}
\vec f_\su{IR} &\eqdef \frac{\rho\kappa_\su{IR}}c\vec F_\su{IR}, \\
\vec f_\su{UV} &\eqdef \frac{\rho\kappa_\su{UV}}c\vec F_\su{UV}.
\end{align}
Note that $\vec f_\su{adv}$ contains $\vec f_\su{cent}$. If we restrict
ourselves to dynamics in the poloidal plane, the difference $\vec
f_\su{adv}-\vec f_\su{cent}$ can be understood in two equivalent ways: either
as the rate of change of local momentum due to poloidal advection, or as the
force $\vec f_\su{ram}$ arising from ram pressure. The individual forces
combine to form the Eulerian and \textquote{Lagrangian} non-gravitational
forces:
\begin{align}
\label{eq:Eulerian force}
\vec f_\su E &\eqdef \vec f_\su{adv}
  +\vec f_\su{gas}+\vec f_\su{mag}+\vec f_\su{IR}+\vec f_\su{UV}, \\
\label{eq:Lagrangian force}
\vec f_\su L &\eqdef \vec f_\su{cent}
  +\vec f_\su{gas}+\vec f_\su{mag}+\vec f_\su{IR}+\vec f_\su{UV}.
\end{align}
The Eulerian non-gravitational force is easy to grasp: If $\vec f_\su E+\vec
f_\su{grav}$ vanishes, then the flow is time-steady, but gas may still
accelerate along streamlines. To interpret the \textquote{Lagrangian}
non-gravitational force, we consider the force equation for a gas packet:
\begin{equation}
\rho\od{\vec v}t=\vec f_\su{grav}
  +\vec f_\su{gas}+\vec f_\su{mag}+\vec f_\su{IR}+\vec f_\su{UV}.
\end{equation}
Since the $R$- and $z$\nobreakdash-components of the left-hand side are $\rho
[\ods{v_R}t-v_\phi (\ods\phi t)]$ and $\rho (\ods{v_z}t)$ respectively, the
same components of $\vec f_\su L+\vec f_\su{grav}$ are $\rho (\ods{v_R}t)$ and
$\rho (\ods{v_z}t)$. The \textquote{Lagrangian} non-gravitational force
therefore tells us how the gas packet moves in the poloidal plane; the special
case of $\vec f_\su L+\vec f_\su{grav}$ having zero poloidal projection means
that the gas packet does not move poloidally.

\Cref{fig:support} compares non-gravitational against gravitational forces in
radial and vertical directions. Only forces that explain support are included:
The radial components of $\vec f_\su{gas}$ and $\vec f_\su{mag}$ are not shown
because they, ignoring signs, are \num{\lesssim0.1} and \num{\lesssim0.05}
times gravity in the body; similarly, the vertical component of $\vec
f_\su{mag}$ is omitted due to it being on average weaker than most other
forces. Although the torus is asymmetric about the mid-plane throughout the
simulation, it must on average be symmetric in the long run; therefore, we fold
each quantity in the figure about the mid-plane by averaging the quantity with
its vertical reflection. We then smooth out fluctuations by averaging over the
interval $6\,t_0\le t\le 14\,t_0$ in which the torus is in \textquote{steady}
state. The figure also shows density contours: The head and the body are the
vertically extended and flat structures about the mid-plane at
$0.5\,r_0\lesssim R\lesssim1.8\,r_0$ and $1.8\,r_0\lesssim R\lesssim3.9\,r_0$
respectively, whereas the wings refer to the region directly above them at
$\tau_\su{UV}\gtrsim1$.

\begin{figure*}
\ifaastex{\vskip-13.5ex}\relax
\includegraphics{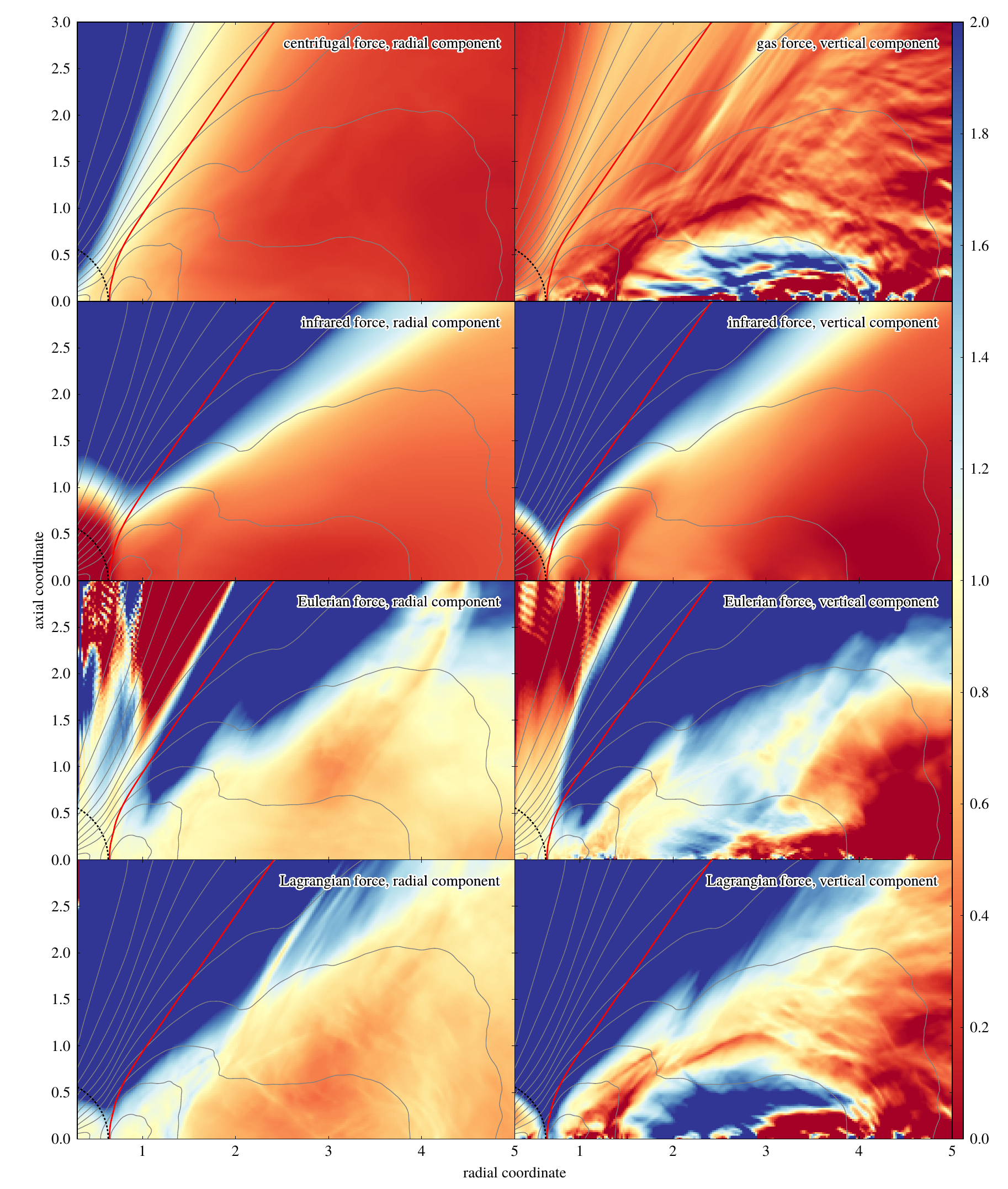}
\caption{Ratios of various forces to gravity in the zoom-in of the
time-averaged, azimuthally averaged, and vertically folded poloidal plane in
the \textquote{steady} state of the \irrmhd{} stage. Colors represent $\uvec
n\cdot\vec f/(-\uvec n\cdot\vec f_\su{grav})$, where $\uvec n\eqdef\uvec e_R$
and $\uvec n\eqdef\uvec e_z$ in the left and right columns; blue and red mean
$\vec f$ provides support stronger and weaker than gravity respectively. The
label in the top-right corner of each panel indicates the value of $\vec f$, as
defined in \cref{eq:centrifugal force,eq:pressure force,eq:IR force,eq:Eulerian
force,eq:Lagrangian force}. The dust sublimation surface $r=r_\su{ds}$
(\cref{eq:dust sublimation radius}) is the dotted black contour around the
origin, the red contour traces the surface on which $\tau_\su{UV}=1$, and gray
contours plot $\rho/\rho_0$ from \numrange{e-5}{1} in logarithmic steps of
\num{e0.5}. All quantities are normalized to fiducial units
(\cref{tab:fiducial}).}
\label{fig:support}
\end{figure*}

Gravity is closely matched both radially and vertically by $\vec f_\su E$ in a
sizable region that encompasses the body and the densest parts of the head.
This implies the velocity field in the region changes little over time, so
structures there have time-steady morphologies, which is what we claimed in
\cref{sec:steady state}. All forces contribute to create the approximate
equality between $\vec f_\su E$ and gravity, with the relative significance of
each force varying by position. The head and the body are inflated in the
vertical direction by both gas and \ac{IR} radiation pressure; the same can be
said of the lower parts of the wings directly above the body, in the triangular
region bounded by $2\,r_0\lesssim R\lesssim3\,r_0$, $\abs z\gtrsim0.5\,r_0$,
and $\abs z/R\lesssim0.7$. The head appears taller than the body because
\ac{IR} and \ac{UV} radiation push the inner edge up into a wind. \Ac{UV} and
\Ac{IR} radiation are the primary drivers of outflow in the wind and in the
upper parts of the wings at $\tau_\su{UV}\gtrsim1$ respectively. These are
exactly the same observations we made about our previous simulations
\citepalias{2016ApJ...825...67C}, but the tori in those simulations were never
in a \textquote{steady} state long enough for concrete remarks to be made.

In the radial direction, $\vec f_\su L$ only partially supports the body
against gravity, so gas accelerates radially as it falls inward through the
body. In the vertical direction,
\begin{equation}
(\sign z)\,\uvec e_z\cdot(\vec f_\su E-\vec f_\su L)
  =(\sign z)\,\uvec e_z\cdot\vec f_\su{ram}<0;
\end{equation}
we interpret this as advection bringing gas with downward momentum to the body,
or equivalently, as vertically collapsing gas exerting downward ram pressure on
the body. The origin of this collapsing gas will be discussed in
\cref{sec:streamlines}. Ram pressure squeezes the body vertically, making the
vertical extent of the body smaller than its sound speed would otherwise
suggest.

The vertical components of $\vec f_\su{adv}$, $\vec f_\su{gas}$, $\vec
f_\su{mag}$, $\vec f_\su{IR}$, and $\vec f_\su{UV}$ in the head, normalized by
the negative of the vertical component of $\vec f_\su{grav}$, are
\num{\approx-0.01}, \num{\approx0.42}, \num{\approx0.02}, \num{\approx0.44},
and \num{\approx0.18} respectively; analogous quantities in the body are
\num{\approx-0.17}, \num{\approx0.67}, \num{\approx0.15}, \num{\approx0.35},
and \num{\approx0.00}. Clearly vertical force balance prevails in the body, so
gas motion is nearly horizontal; moreover, gas pressure is chiefly responsible
for counteracting downward forces. We shall extrapolate these results to
realistic \acp{AGN} in \cref{sec:extrapolation}.

\subsection{Streamlines}
\label{sec:streamlines}

\Cref{fig:streamlines} depicts streamlines in regions of \textquote{steady}
state, where the flow timescale is shorter than the simulation duration. The
global inflow--outflow is now manifest. The head, body, wind, and wings are not
hydrostatic; rather, gas from, say, $3\,r_0\lesssim R\lesssim4\,r_0$ and
$r_0\lesssim\abs z\lesssim2\,r_0$ migrates first to the body, then to the head,
and finally to the wind or the wings. The four structures are recognizable
during the entire \textquote{steady} state because they retain qualitatively
similar shapes (\cref{sec:steady state}) even as gas passes through them.

\begin{figure*}[!t]
\includegraphics{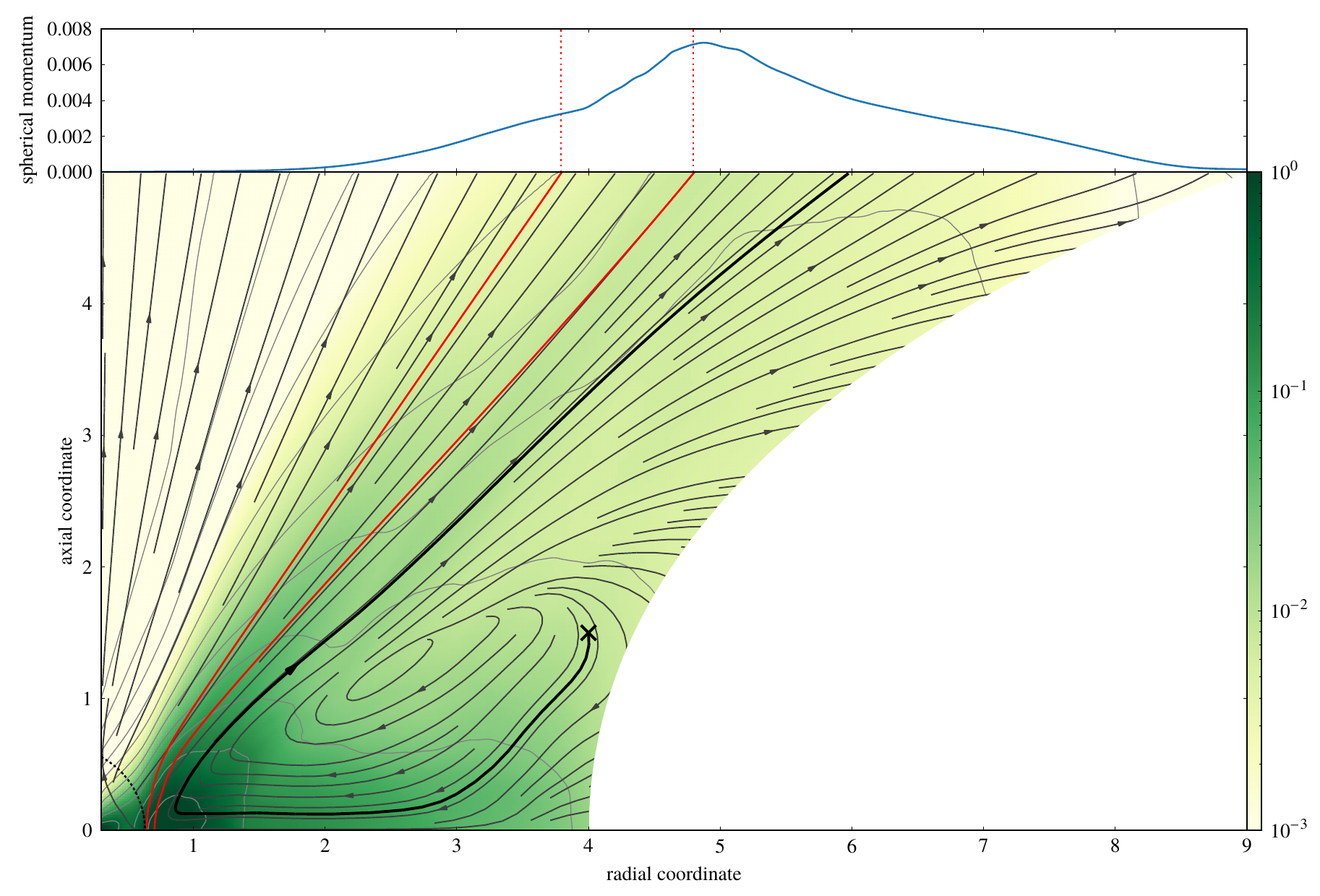}
\caption{Gas flow in the time-averaged, azimuthally averaged, and vertically
folded poloidal plane in the \textquote{steady} state of the \irrmhd{} stage.
All quantities are normalized to fiducial units (\cref{tab:fiducial}).
\textit{Top panel:} Plot of $\uvec e_r\cdot(\rho\vec v)$ along the vertical
boundaries of the simulation domain. The left and right vertical dotted lines
are drawn at $\tau_\su{UV}=1$ and $\tau_\su{UV}=4$. \textit{Bottom panel:}
Colors represent $\rho\norm{\vec v}$, and dark gray streamlines follow gas
velocity. The marked streamline is discussed in \cref{sec:streamlines}. The
dust sublimation surface $r=r_\su{ds}$ (\cref{eq:dust sublimation radius}) is
the dotted black contour around the origin, red contours trace the surfaces on
which $\tau_\su{UV}=1$ and $\tau_\su{UV}=4$, and light gray contours plot
$\rho/\rho_0$ from \numrange{e-5}{1} in logarithmic steps of \num{e0.5}.}
\label{fig:streamlines}
\end{figure*}

Let us follow the marked streamline in \cref{fig:streamlines}. Gas and \ac{IR}
radiation together furnish less radial and vertical support against gravity in
the \irrmhd{} stage than gas with artificially elevated pressure (\cref{sec:MHD
stage}) did in the \uvrmhd{} stage; indeed, \cref{fig:support} shows $\vec
f_\su L$ to be weaker than gravity in both directions at the starting point of
the streamline. Weaker support, combined with the fact that gas was already
infalling in the \uvrmhd{} stage (\cref{sec:UV dynamics}), means that gas
streams inward radially and vertically, as demonstrated by the bottom row of
\cref{fig:overview} and by \cref{fig:streamlines}. This gas piles onto the
upper surface of the body, generating ram pressure (\cref{sec:forces}). The
reaction by the body on the gas results in a shock, visible in
\cref{fig:overview}, which removes the vertical component of velocity from the
gas; this is why $\vec f_\su L$ in \cref{fig:support} points strongly upward
throughout the body.

Gas in the body moves horizontally inward because $\vec f_\su E$ nearly
balances gravity in the vertical direction (\cref{sec:forces}). This journey is
arrested when gas reaches the irregular surface separating the head from the
body. We witness in \cref{fig:overview} that a shock along this interface
strips the gas of much of its remaining radial component of velocity. Flow in
the head is much slower than in the body, but \cref{fig:streamlines} reveals
that gas on the whole is still moving horizontally inward to the inner edge. As
soon as it gets there, it is ejected by \ac{UV} radiation.

\subsection{Wide-angle outflow}
\label{sec:outflow}

The streamlines in \cref{fig:streamlines} suggest that while part of the gas
leaving the inner edge becomes the wind at $\tau_\su{UV}\lesssim1$, part of it
recedes to $\tau_\su{UV}\gtrsim1$ and blends into the wings. Outflow occurs
through both structures; what makes the wings special is that they stretch
across a broader latitude range than the wind, and they also host a more
massive outflow: The \textquote{steady}-state mass, momentum, and kinetic
energy loss rates at $\tau_\su{UV}>1$ are respectively \num{\sim6.1},
\num{\sim1.8}, and \num{\sim0.85} times those at $\tau_\su{UV}<1$.

\Ac{IR} radiation diffusing outward from the central hole through the torus is
vital for driving this wide-angle outflow. One piece of evidence supporting
this claim is in \cref{fig:support}: The upper parts of the wings at
$\tau_\su{UV}\gtrsim1$ and $\abs z/R\gtrsim0.7$ experience outward \ac{IR}
radiative acceleration much stronger than gravity. Another is in the top panel
of \cref{fig:streamlines}, which plots the spherically radial component of gas
momentum along the vertical boundaries of the simulation domain. The curve
peaks at $\tau_\su{UV}\approx4.87$, close to
$\tau_\su{UV}=\kappa_\su{UV}/\kappa_\su{IR}=4$, the surface of unit \ac{IR}
optical depth from the origin.

It should be remembered that this division of the outflow into wind and wings
is only for our cognitive convenience. The curve in the top panel of
\cref{fig:streamlines} has no perceptible discontinuity at $\tau_\su{UV}\sim1$;
\ac{IR} and \ac{UV} radiation work in tandem to power a continuous outflow
across a large solid angle, from the wind at $\tau_\su{UV}\lesssim1$ to the
wings at $\tau_\su{UV}\gtrsim1$.

The upper parts of the wings in \cref{fig:overview} appear to be
gravitationally unbound. Given our limited simulation domain, we cannot say
definitively if the outflow in the wings would reach infinity. If the infalling
part of the torus has a flaring shape, as it ostensibly does in
\cref{fig:overview}, and if gas velocity in the wings does not make a large
enough angle with the mid-plane, then the outflow may eventually run into the
inflow. The outflow may lose energy in shocks, become bound, and merge with the
inflow, thus creating a circulation of gas in the torus.

Wings are the direct analogues of palisades in the \uvrmhd{} stage
(\cref{sec:UV dynamics}). Much that applies to palisades carries over to wings:
Both are similarly located in the torus, both have gas flowing outward only
above a certain latitude, and both contain small-scale density inhomogeneities
caused by bursty wind launching and consequent shocks. The two structures are
nevertheless very different dynamically: Palisades are not propelled by \ac{IR}
radiation, move much more slowly, and are gravitationally bound.

We saw in \cref{sec:UV dynamics} how the angle between the palisades and the
mid-plane can be estimated from just three parameters; here we carry out the
same analysis \foreign{mutatis mutandis}. While the gas-supported torus in the
\uvrmhd{} stage had $c_\su s^2\sim GM/r_\su{ds}$ (\cref{sec:UV dynamics}), the
torus in the \irrmhd{} stage has $c_\su s^2\sim R_\su{ideal}T_\su{ds}\ll
GM/r_\su{ds}$. It is obvious from \cref{fig:palisade} that reducing $c_\su
s/v_\infty$ brings $\psi$ closer to $\chi$; in the limit of $c_\su
s/v_\infty\lesssim0.1$, as is applicable to our simulation, \cref{eq:wing
angle} yields $\psi-\chi\approx(c_\su s/v_\infty)\sin\chi$. This means wings
are almost parallel to the inner surface, in keeping with \cref{fig:overview}.

\subsection{\texorpdfstring{\acs*{IR}}{IR} radiation}
\label{sec:IR}

We define the normalized \ac{IR} radiative flux and radiation energy density as
$[L_\su{UV}/(4\pi r^2)]^{-1}\,\uvec e_r\cdot\vec F_\su{IR}$ and
$[L_\su{UV}/(4\pi r^2)]^{-1}\,(cE_\su{IR})$ respectively; the former quantity
is unity if all \ac{UV} radiation were converted to the \ac{IR}, and if \ac{IR}
radiative flux were spherically symmetric. \Cref{fig:IR angular moments} plots
the quantities measured on two surfaces: Blue curves are for a sphere of radius
$r=3\,r_0$ cutting through the body; orange curves are for the outer-radial and
vertical boundaries of the simulation domain. Our simulation domain is large
enough for the normalized \ac{IR} radiative flux to asymptote at large
distances. The wiggles along the curves are the consequence of the \ac{IR}
\ac{RT} module directing radiation into preferred directions in optically thin
regions (\cref{sec:radiative transfer}) and are not physical. The wiggles do
not die out with distance, but fluctuations about the mean are at worst
\SI{\lesssim20}{\percent}.

\begin{figure}
\includegraphics{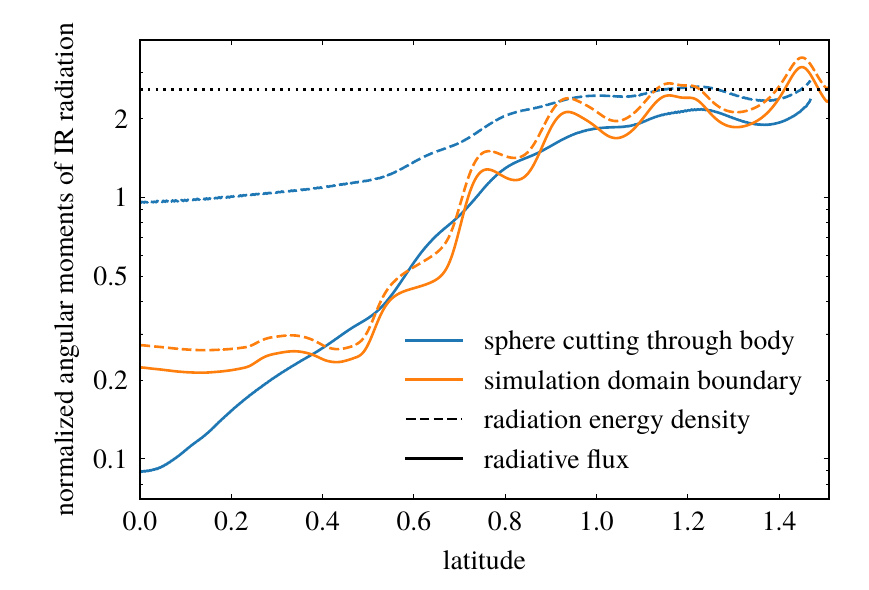}
\caption{Plot of time-averaged, azimuthally averaged, and vertically folded
normalized angular moments of \ac{IR} radiation against latitude in the
\textquote{steady} state of the \irrmhd{} stage. The solid and dashed curves of
each color are for $[L_\su{UV}/(4\pi r^2)]^{-1}\,\uvec e_r\cdot\vec F_\su{IR}$
and $[L_\su{UV}/(4\pi r^2)]^{-1}\,(cE_\su{IR})$ respectively. Blue curves are
for $r=3\,r_0$; orange curves are for the outer-radial and vertical boundaries
of the simulation domain. The horizontal dotted line is drawn at
$C_\su{IR}/(1-C_\su{IR})$, where $C_\su{IR}$ is the time-averaged \ac{IR}
covering fraction (\cref{sec:obscuration}).}
\label{fig:IR angular moments}
\end{figure}

First consider the blue curves. The solid curve is \num{\lesssim0.1} times the
dashed curve at low latitudes; this means \ac{IR} radiation is fairly isotropic
inside the body. In addition, the value of the solid curve at low latitudes is
\num{\lesssim0.05} times that at high latitudes; this is indicative of the
degree to which the optically thick torus concentrates \ac{IR} radiative flux
into the polar direction \citepalias{2016ApJ...825...67C}.

Next we compare the solid curves. The curves almost coincide at latitudes
\num{\gtrsim0.4}; this suggests $\uvec e_r\cdot\vec F_\su{IR}\propto r^{-2}$ in
the central hole, the wind, and the wings, from as small a radius as $r=3r_0$
outward. Both curves approach $\mathrelp\sim C_\su{IR}/(1-C_\su{IR})$ near the
axis, in agreement with our previous result \citepalias{2016ApJ...825...67C};
here $C_\su{IR}$ is the \ac{IR} covering fraction, defined as the fraction of
sightlines toward the central source with \ac{IR} optical depth above unity.
The blue curve falls below the orange at latitudes \num{\lesssim0.4}, but this
is simply because the head and the body have limited radial extent, so outgoing
\ac{IR} radiation can diffuse around and reach regions behind them.

\subsection{Temperature}
\label{sec:temperature}

Gas and \ac{IR} radiation temperature contours in \cref{fig:temperature} are
very close to each other at $\tau_\su{UV}\gtrsim1$. They are not spherical
because the dashed curves in \cref{fig:IR angular moments} increase with
latitude; in our particular simulation, the contours are strikingly vertical
from the mid-plane almost up to the $\tau_\su{UV}=1$ surface. The dashed orange
curve in \cref{fig:IR angular moments} has greater variation over latitude than
the blue; accordingly, contours further from the origin are less spherical.

\begin{figure}
\includegraphics{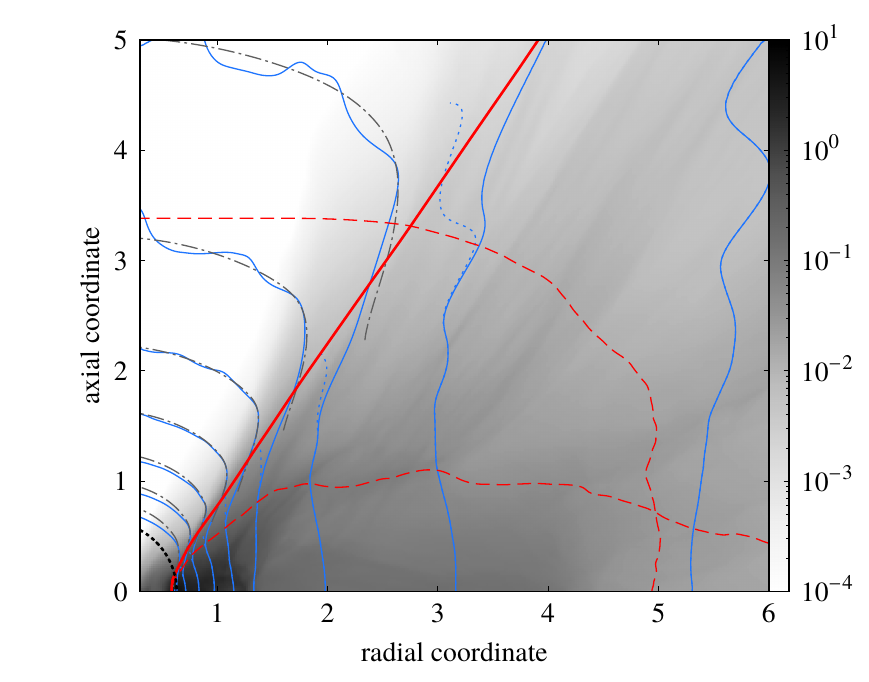}
\caption{Zoom-in of the azimuthally averaged and vertically folded poloidal
plane in the \irrmhd{} stage at $t=14\,t_0$. Solid and dotted blue contours
show gas and \ac{IR} radiation temperatures; dash-dotted gray contours show the
modeled temperature given by \cref{eq:central hole temperature}. Temperatures
on these contours go down from $T_\su{ds}$ in steps of $0.1\,T_\su{ds}$ as one
moves away from the origin. Only parts of the \ac{IR} and modeled temperature
contours are drawn to reduce clutter. The upper and lower dashed red contours
are the edge-on and face-on \ac{IR} photospheres (\cref{sec:photosphere}). Gas
density is presented on a logarithmic scale as gray intensities (see color bar
along the right edge), the dust sublimation surface $r=r_\su{ds}$
(\cref{eq:dust sublimation radius}) is the dotted black contour around the
origin, and the solid red contour traces the surface on which $\tau_\su{UV}=1$.
All quantities are normalized to fiducial units (\cref{tab:fiducial}).}
\label{fig:temperature}
\end{figure}

Temperature in the central hole has a spherically symmetric distribution, with
deviations only at the angles into which the \ac{IR} \ac{RT} module
concentrates \ac{IR} radiation (\cref{sec:radiative transfer}). It can be
modeled by considering the balance of energy emitted and absorbed by a dust
grain of radius $a$ in thermal equilibrium:
\begin{equation}
4\pi a^2\kappa_\su{IR}\frac{c\aSB}4T^4=
  \pi a^2\kappa_\su{UV}\frac{L_\su{UV}}{4\pi r^2}e^{-\tau_\su{UV}}+
  \pi a^2\kappa_\su{IR}F_\su{IR}.
\end{equation}
Following \citetalias{2016ApJ...825...67C}, the \ac{IR} radiative flux is
$F_\su{IR}\approx C_\su{UV}/(1-C_\su{IR})\times L_\su{UV}/(4\pi r^2)$. The
\ac{UV} absorption term contains the \ac{UV} radiative flux corrected for
extinction; such correction is unnecessary for the \ac{IR} because
$\tau_\su{UV}\lesssim1$ automatically implies $\tau_\su{IR}\ll1$, where
$\tau_\su{IR}$ is the \ac{IR} optical depth from the central source.
Rearranging, we get
\begin{equation}\label{eq:central hole temperature}
\aSB T^4\approx\frac{L_\su{UV}}{4\pi r^2c}
  \left(\frac{\kappa_\su{UV}}{\kappa_\su{IR}}e^{-\tau_\su{UV}}
  +\frac{C_\su{UV}}{1-C_\su{IR}}\right).
\end{equation}
Using the operational definition of $C_\su{IR}$ and the simplification
$C_\su{UV}\approx C_\su{IR}$ from \citetalias{2016ApJ...825...67C}, we plot the
modeled temperature in \cref{fig:temperature}; the model is excellent at
$\tau_\su{UV}\lesssim1$. \Cref{eq:opacity temperature,eq:central hole
temperature} agree if $\tau_\su{UV}\ll1$ and $F_\su{IR}=0$, hence the opacity
temperature in the \uvrmhd{} stage is consistent with the actual temperature in
the \irrmhd{} stage.

\subsection{Magnetic field}
\label{sec:magnetic field}

\Cref{fig:plasma beta} graphs gas-only plasma beta $\beta_\su g\eqdef
p/(\tfrac12B^2)$ and total plasma beta $\beta_\su
t\eqdef(p+\frac13E_\su{IR})/(\tfrac12B^2)$ at $t=0$ and $t=14\,t_0$. The range
of $\beta_\su g$ varies little across space and time. The same can almost be
said of $\beta_\su t$, except that its value at $t=14\,t_0$ is much higher in
the central hole and the upper parts of the wings than the rest of the
simulation domain. This is because $E_\su{IR}$ is a few times higher in the
central hole than in other parts of the torus (\cref{sec:IR}), while $p$ is
several orders of magnitude lower; it then follows from the definitions of the
plasma betas that $\beta_\su t\gg\beta_\su g$. The spatial distributions of
$\beta_\su g$ and $\beta_\su t$ are virtually identical apart from the overall
normalization, and apart from the central hole. The temporal constancy of
$\beta_\su g$ and $\beta_\su t$ means that complex gas motion does not
perceptibly modify the \ac{MHD} saturation state, at least not within our
finite simulation time. Additionally, \cref{fig:plasma beta} suggests that only
the wings, where outflow drags out field lines, have large-scale order in the
magnetic field, but even there the field does not point uniformly inward or
outward, and neighboring regions can have fields in opposite directions.

\begin{figure*}
\includegraphics{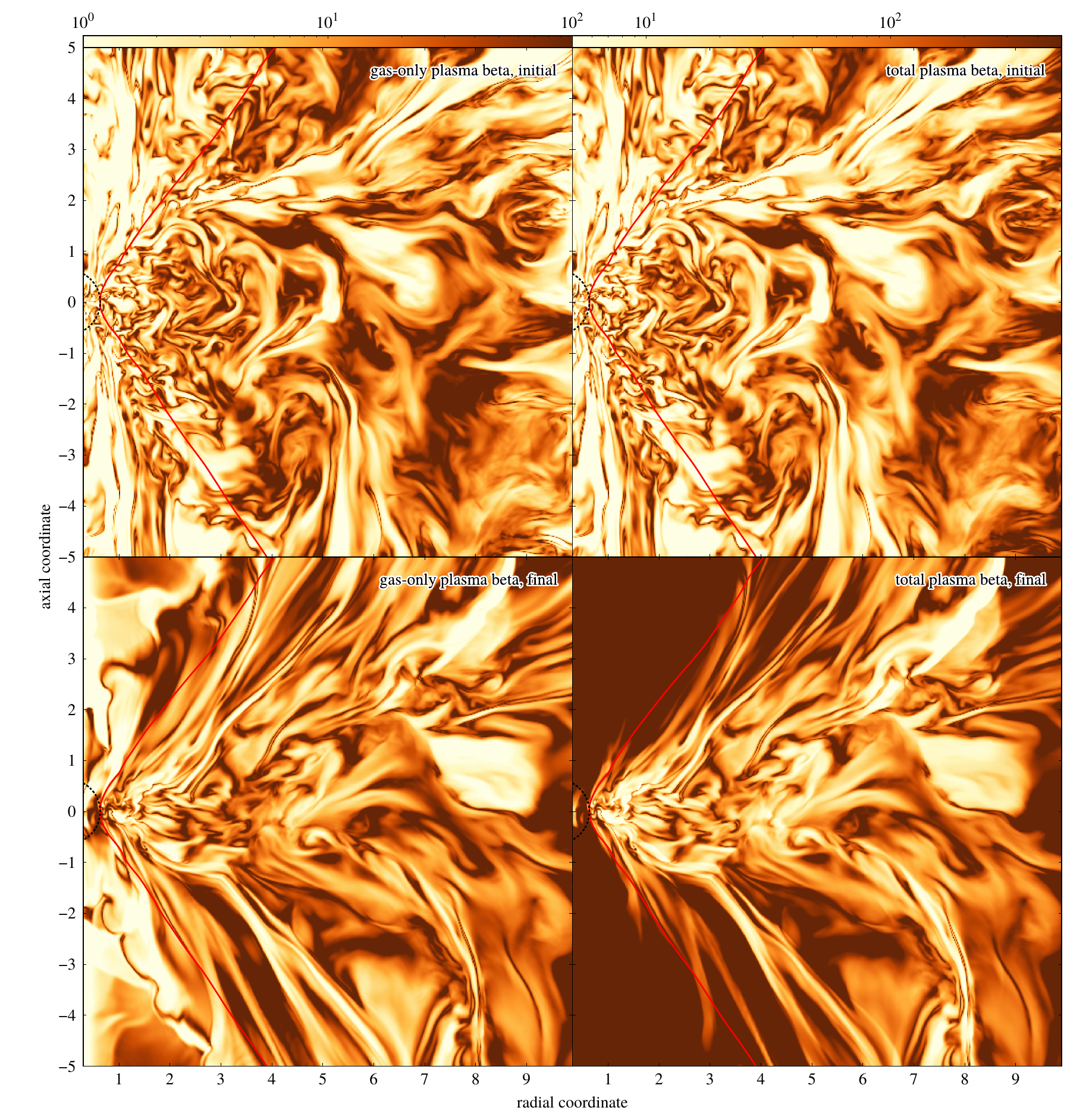}
\caption{Plasma betas in the poloidal plane along $\phi=0$ in the \irrmhd{}
stage. Colors in the left and right columns represent gas-only and total plasma
beta respectively (\cref{sec:magnetic field}; see color bars along the top
edge). The top and bottom rows are at $t=0$ and $t=14\,t_0$. The dust
sublimation surface $r=r_\su{ds}$ (\cref{eq:dust sublimation radius}) is the
dotted black contour around the origin, and the red contour traces the surface
on which $\tau_\su{UV}=1$. All quantities are normalized to fiducial units
(\cref{tab:fiducial}).}
\label{fig:plasma beta}
\end{figure*}

We define the density-weighted plasma betas and alpha parameter as
\begin{align}
\mean{\beta_\su g} &\eqdef \frac{\int dV\,\rho p}
  {\int dV\,\rho (\tfrac12B^2)}, \\
\mean{\beta_\su t} &\eqdef \frac{\int dV\,\rho (p+\tfrac13E_\su{IR})}
  {\int dV\,\rho (\tfrac12B^2)}, \\
\mean\alpha &\eqdef \frac{\int dV\,\rho (\rho v_R\Delta v_\phi-B_RB_\phi)}
  {\int dV\,\rho (p+\tfrac13E_\su{IR})},
\end{align}
where $\Delta v_\phi$ is the perturbation of $v_\phi$ about its azimuthal
average. We see $\mean{\beta_\su g}$ rise steadily from \num{\sim4.2} to
\num{\sim4.6} over $0\lesssim t\lesssim10\,t_0$ and more rapidly to
\num{\sim7.5} at $t=14\,t_0$, while $\mean{\beta_\su t}$ drops from
\num{\sim27} to \num{\sim19} over $0\lesssim t\lesssim2\,t_0$ and holds still
at \num{\sim16} thereafter. In addition, $\mean\alpha$ climbs from
\num{\sim0.019} to \num{\sim0.044} over $0\le t\le14\,t_0$. The \acp{MHD}
simulations cited by \citet{2011ApJ...738...84H} have $\beta\sim15$ and
$\alpha\sim0.02$, while the \acp{RMHD} simulations by
\citet{2009ApJ...691...16H} have $\alpha\sim0.02$; since our $\mean{\beta_\su
t}$ and $\mean\alpha$ are close to these values, \acp{MHD} turbulence is likely
at saturation.

We see large spatial variation in plasma betas in \cref{fig:plasma beta}
because field fluctuations are of the same order as the mean. The associated
velocity fluctuations, having Mach numbers \num{\sim0.1}, are much smaller than
the global flow speed, so gas motion remains well ordered throughout the
simulation, and the smooth time-averaged streamlines in \cref{fig:streamlines}
closely resemble the flow at any one time.

\section{Discussion}
\label{sec:discussion}

In this section, we consider the observational implications of the \irrmhd{}
stage (\cref{sec:photosphere,sec:obscuration}), extrapolate results from that
stage to parameters applicable to realistic tori (\cref{sec:extrapolation}),
and compare our simulation with other torus models with outflows
(\cref{sec:comparison}).

\subsection{Observed temperature profiles in \titleirrmhd{} stage}
\label{sec:photosphere}

Interferometric observations of tori are analyzed by fitting ellipsoidal
blackbodies of various sizes and temperatures to visibilities; the result can
be interpreted as a crude temperature profile
\citep[e.g.,][]{2007A&A...474..837T}. To facilitate comparison of the
\textquote{steady}-state torus in the \irrmhd{} stage with observations, it is
useful to locate its \ac{IR} photospheres as seen by observers both face-on and
edge-on, and to determine its observed temperature profiles. Here we adopt the
gray-opacity approximation of our simulation, that is,
$\kappa_\su{IR}/\kappaT=20$ wherever $T\lesssim T_\su{ds}$ (\cref{sec:radiative
transfer}). We ignore the possibility that rarefied gas at
$\tau_\su{UV}\lesssim1$ could be much hotter than in our simulation because we
do not treat its physics accurately, and because its Thomson optical depth
along the vertical sightlines described below is \num{\lesssim5e-3}.

The lower dashed red contour in \cref{fig:temperature} shows the face-on
\ac{IR} photosphere of the torus, that is, where the \ac{IR} optical depth as
measured vertically from the upper-vertical boundary equals unity. The
photosphere is approximately horizontal, separating the wings above from the
head and the body below. Vertical sightlines at $R\gtrsim r_\su{ds}$ always
intersect the photosphere; the photospheric temperature has a gradient
$\mathrelp\propto R^{-0.73}$ starting from $\mathrelp\approx T_\su{ds}$ at
$R\approx r_\su{ds}$. The observed \ac{IR} emission includes the contribution
from all gas above the photosphere; however, gas at $\tau_\su{UV}\gtrsim1$ does
not significantly modify the temperature of the \ac{IR} radiation from the
photosphere because temperature contours inside this gas are close to vertical
(\cref{sec:temperature}). According to \cref{fig:temperature}, if we were to
enlarge the simulation domain vertically, the part of the photosphere at
$R\lesssim3\,r_0$ would barely change because sightlines from the
upper-vertical boundary would simply encounter additional amounts of rarefied
gas at $\tau_\su{UV}\lesssim1$, but the part at $R\gtrsim3\,r_0$ would shift
upward noticeably because sightlines would cut through more of the denser gas
at $\tau_\su{UV}\gtrsim1$. This vertical displacement of the photosphere should
not qualitatively change the observed temperature gradient since temperature
contours are vertical (\cref{sec:temperature}).

The upper dashed red contour shows the edge-on \ac{IR} photosphere, that is,
where the \ac{IR} optical depth as measured horizontally from the outer-radial
boundary equals unity. Horizontal sightlines at $\abs z\lesssim2\,r_0$
intersect the vertical portion of the photosphere at $R\sim5\,r_0$; the fact
that this portion has temperature $\mathrelp\sim0.2\,T_\su{ds}$ over its
entirety may be germane to the observed constant temperature of
\SI{\sim300}{\kelvin} at \SI{\lesssim2}{\parsec} scales in Circinus
\citep{2014A&A...563A..82T}. Horizontal sightlines at $2\,r_0\lesssim\abs
z\lesssim3.4\,r_0$ intersect the photosphere where its tangent $\abs{\ods Rz}$
is large, so the observed temperature on these sightlines rises rapidly with
$\abs z$; however, this part of the photosphere presents only a small projected
area to edge-on views. Finally, horizontal sightlines at $\abs
z\gtrsim3.4\,r_0$ do not intersect the photosphere at all; the \ac{IR} emission
along these sightlines is likely weaker because they are optically thin, but
hotter because they pass through gas with higher average temperature than
sightlines at $\abs z\lesssim3.4\,r_0$. If we were to expand the simulation
domain radially, the photosphere would necessarily move outward, but the
magnitude of the shift depends on the unknown gas distribution outside the
simulation domain. Nevertheless, since density generally falls off with height,
the presence of additional opacity at greater radii should only boost
$\abs{\ods Rz}$ of the photosphere, so we would still expect a sharp increase
in observed temperature at some $\abs z$.

\subsection{\xray, \texorpdfstring{\acs*{IR}}{IR}, and
\texorpdfstring{\acs*{UV}}{UV} obscuration in \titleirrmhd{} stage}
\label{sec:obscuration}

The left panel of \cref{fig:optical depths} plots Thomson, \ac{IR}, and \ac{UV}
optical depths against latitude for one snapshot in the \irrmhd{} stage. To fit
the curves into the same scale, \ac{IR} and \ac{UV} optical depths have been
divided by $\bar\kappa_\su{IR}/\kappaT=20$ and $\bar\kappa_\su{UV}/\kappaT=80$
respectively (\cref{sec:radiative transfer}); the actual optical depths can be
discerned by means of the dotted lines marking the normalized optical depths at
which sightlines become optically thick. The \ac{IR} and \ac{UV} curves are
indistinguishable because $\kappa_\su{IR}$ and $\kappa_\su{UV}$ have identical
temperature dependence; the two curves are very close to the Thomson curve
because $\kappa_\su{IR}$ and $\kappa_\su{UV}$ depend weakly on temperature in
most of the torus (\cref{sec:radiative transfer}). The thinness of the shaded
regions establishes that optical depth in our torus is largely independent of
azimuth.

\begin{figure*}
\includegraphics{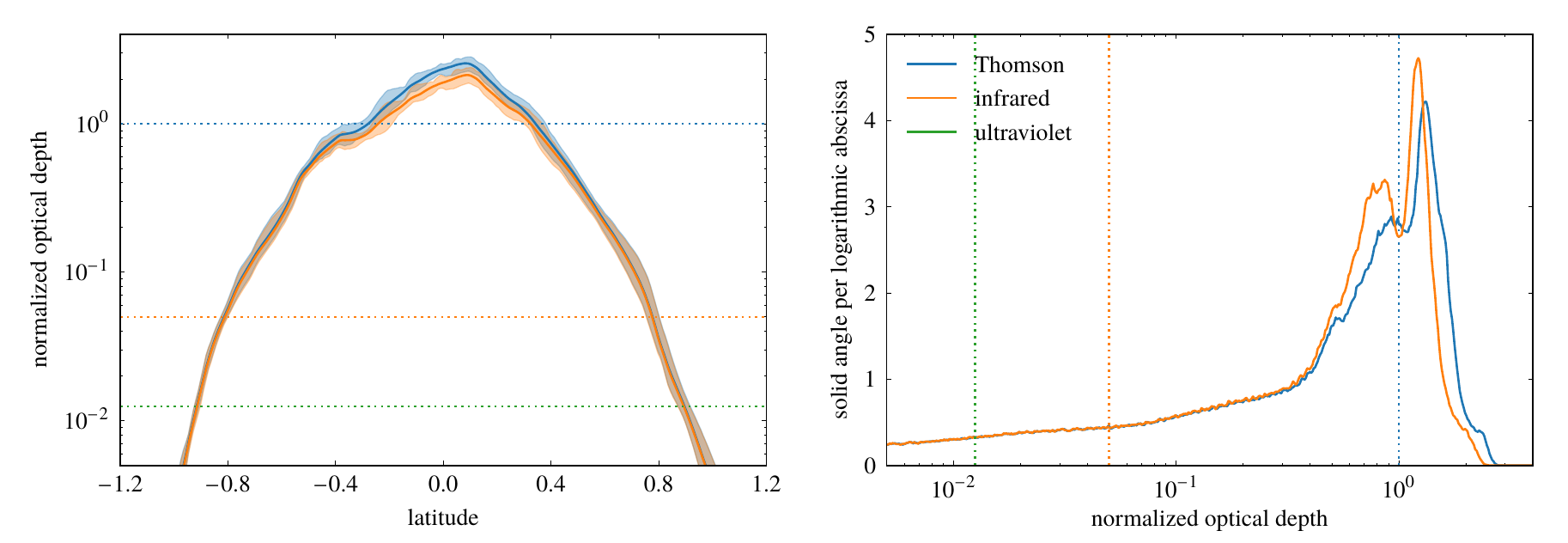}
\caption{\textit{Left panel:} Plot of normalized optical depths
(\cref{sec:obscuration}) along sightlines of varying latitudes in the \irrmhd{}
stage at $t=14\,t_0$. The thick curve is the azimuthal average, and the shaded
area is the range covered by all azimuthal angles. \textit{Right panel:}
Histograms of the solid angle around the origin occupied by gas columns with
normalized optical depths in each logarithmic bin in the \textquote{steady}
state of the \irrmhd{} stage. \textit{Both panels:} Curves for \ac{IR} and
\ac{UV} are identical. Dotted lines indicate optical depth of unity at each
frequency.}
\label{fig:optical depths}
\end{figure*}

The right panel plots $\ods\Omega{\log_{10}\tau}$, the solid-angle coverage
around the origin as a function of logarithmic normalized optical depths,
during the entire \textquote{steady} state of the \irrmhd{} stage. The
similarity of the histograms is again due to the weak temperature dependence of
$\kappaT$, $\kappa_\su{IR}$, and $\kappa_\su{UV}$. The peak at larger optical
depth corresponds to sightlines through both the head and the body, while the
peak at smaller optical depth is due to sightlines through the head only.

We define obscuration in \xrays{} by $\tau_\su T=0.01$: Neutral gas of such
column density is optically thick to \xrays{} at
\SI{\lesssim2}{\kilo\electronvolt}. Likewise, we define obscuration in the
\ac{IR} and \ac{UV} by $\tau_\su{IR}=1$ and $\tau_\su{UV}=1$ respectively. For
our assumed mid-plane Thomson optical depth of \num{\sim2} (\cref{sec:IR-RMHD
stage}), the soft \xray, \ac{IR}, and \ac{UV} covering fractions in the
\textquote{steady} state of the \irrmhd{} stage fluctuate within the ranges
$0.78\lesssim C_\su{soft}\lesssim0.83$, $0.71\lesssim C_\su{IR}\lesssim0.73$,
and $0.77\lesssim C_\su{UV}\lesssim0.82$ respectively; these ranges are close
to one another, and also to the observed fraction of type\nobreakdash-2
\acp{AGN} \citep[e.g.,][]{2010ApJ...714..561L}. Such broad coverage of the
central source is achieved by combining the head, body, wings, and wind. In
contrast, hard \xrays{} up to a few hundred \si{\kilo\electronvolt} are blocked
along Compton-thick sightlines with $N_\element
H\gtrsim\SI{e24}{\per\centi\meter\squared}$ or $\tau_\su T\ge1$; only
sightlines traversing both head and body have such high optical depths, so they
are concentrated near the mid-plane, taking up a solid angle of $0.15\lesssim
C_\su{hard}\lesssim0.28$. This covering fraction coincides with the fraction of
Compton-thick \acp{AGN}, observed to be \numrange{\sim0.2}{\sim0.3}
\citep[e.g.,][]{2015ApJ...815L..13R, 2016ApJ...825...85K}.

Density profiles along sightlines to the central source serve as diagnostics of
the statistics of density fluctuations, sometimes called \textquote{clumping}
in this context. We graph these profiles in \cref{fig:density}, leaving out
regions not in a \textquote{steady} state (\cref{sec:transient}). Apart from
their overall scales, all profiles are similar in that density along a
sightline decreases outward, but with fluctuations. These fluctuations are a
blend of turbulent structures and, for latitudes \SI{\gtrsim0.4}{\radian},
density ridges in the wings (\cref{sec:steady state}); their density contrast
goes from \SI{\lesssim50}{\percent} at low latitudes up to a factor of a few at
high latitudes. These fluctuations, with their irregular shapes and small
amplitudes, do not at all resemble the symmetrical, isolated clumps posited in
many phenomenological \ac{RT} models \citep[e.g.,][]{2002ApJ...570L...9N,
2006A&A...452..459H, 2008ApJ...685..147N, 2008A&A...482...67S,
2012ApJ...751...27H, 2012ApJ...759...36R, 2012MNRAS.420.2756S}. In addition, on
sightlines with latitudes \SI{\gtrsim0.4}{\radian}, the spacing between maxima
stretches with latitude because the wings are parallel to the inner surface
(\cref{sec:outflow}).

\begin{figure}
\includegraphics{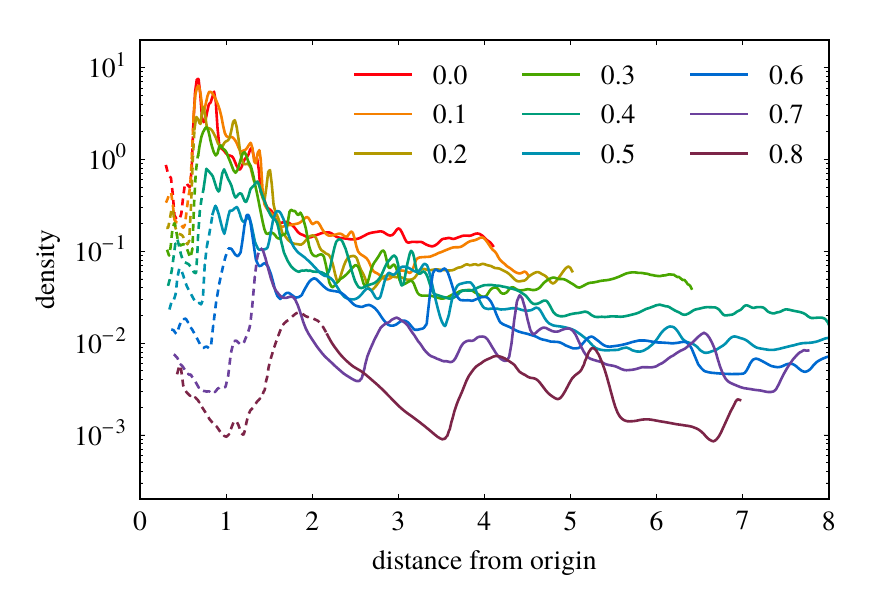}
\caption{Plot of gas density in the \irrmhd{} stage at $t=14\,t_0$ along
sightlines to the central source at fixed azimuth $\phi=0$ and varying
latitudes as indicated in the legend. The dashed and solid portions of each
curve are at $\tau_\su{UV}<1$ and $\tau_\su{UV}>1$ respectively. All quantities
are normalized to fiducial units (\cref{tab:fiducial}).}
\label{fig:density}
\end{figure}

\subsection{Extrapolating to realistic \texorpdfstring{\acs*{AGN}}{AGN} tori}
\label{sec:extrapolation}

Our simulations cannot employ parameters that apply to realistic tori due to
numerical reasons; instead, we must adopt a smaller central mass $M$, or
equivalently, a higher $c_\su s/v_\phi$ \citepalias{2016ApJ...825...67C}, as
well as a lower \ac{UV} opacity $\kappa_\su{UV}$ (\cref{sec:radiative
transfer}). Moreover, our simulations consider just a few values for the
Thomson optical depth $\tau_\su T$, and completely ignore photoionization and
Compton heating. It is imperative that we understand how our results may change
with $c_\su s/v_\phi$, $\kappa_\su{UV}$, and $\tau_\su T$, and how they may be
modified by photoionization and Compton heating.

\subsubsection{Mass loss rate and outflow speed}
\label{sec:realistic outflow}

Mass, momentum, and kinetic energy loss rates are primary observables of our
simulations. To extrapolate loss rates from our simulations to realistic tori,
we introduced in \citetalias{2016ApJ...825...67C} a simple analytic model of an
outflow of unit \ac{UV} optical depth powered by \ac{UV} radiation pressure.
The fiducial mass loss rate and outflow speed of the model are given by
Equations~(25) and~(26) in that article, which we duplicate below:
\begin{align}
\label{eq:UV mass loss rate}
\dot M &\sim 4\pi
  \biggl(\frac{GMR_\su{in}}{\kappaT^2}\frac{L_\su{UV}}{L_\su E}\biggr)^{1/2}
  \biggl(\frac{\kappa_\su{UV}}\kappaT\biggr)^{-1/2}, \\
\label{eq:UV outflow speed}
v_\infty &\sim \biggl(\frac{GM}{R_\su{in}}\frac{L_\su{UV}}{L_\su E}
  \frac{\kappa_\su{UV}}\kappaT\biggr)^{1/2}.
\end{align}
We carry this model over to our current simulations. The model strictly
pertains only to the \ac{UV}-driven outflow in the \uvrmhd{} stage, but it can
be adapted to describe the \ac{IR}-driven part of the outflow in the \irrmhd{}
stage because that part also penetrates to unit \ac{IR} optical depth
(\cref{sec:outflow}). If we multiply $L_\su{UV}$ by $\mathrelp\sim
C_\su{IR}/(1-C_\su{IR})$ in \cref{eq:UV mass loss rate,eq:UV outflow speed} to
account for repeated scattering of \ac{IR} radiation inside the central hole
\citepalias{2016ApJ...825...67C} and replace $\kappa_\su{UV}$ by
$\kappa_\su{IR}$, we find
\begin{align}
\label{eq:IR mass loss rate}
\dot M &\sim 4\pi
  \biggl(\frac{GMR_\su{in}}{\kappaT^2}\frac{L_\su{UV}}{L_\su E}
  \frac{C_\su{IR}}{1-C_\su{IR}}\biggr)^{1/2}
  \biggl(\frac{\kappa_\su{IR}}\kappaT\biggr)^{-1/2}, \\
\label{eq:IR outflow speed}
v_\infty &\sim \biggl(\frac{GM}{R_\su{in}}\frac{L_\su{UV}}{L_\su E}
  \frac{C_\su{IR}}{1-C_\su{IR}}\frac{\kappa_\su{IR}}\kappaT\biggr)^{1/2}.
\end{align}
The $M$- and $\kappa_\su{UV}$\nobreakdash-dependences of \cref{eq:UV mass loss
rate,eq:UV outflow speed,eq:IR mass loss rate,eq:IR outflow speed} form the
basis of our extrapolation. In the remainder of this discussion, we take
$R_\su{in}=r_\su{ds}$ with $r_\su{ds}$ from \cref{eq:dust sublimation radius},
and set $C_\su{IR}\approx0.72$ as found in the \textquote{steady} state of the
\irrmhd{} stage (\cref{sec:obscuration}).

For the \textquote{steady} state of the \uvrmhd{} stage, the mass loss rate
through the vertical boundaries is measured directly, while the outflow speed
is obtained by comparing momentum and kinetic energy loss rates with the mass
loss rate; the two values are then scaled by \cref{eq:UV mass loss rate,eq:UV
outflow speed}. We obtain
\ifaastex{
  \begin{equation}
  \num{\sim0.15}\,
    \biggl(\frac M{\SI{e7}{\solarmass}}\biggr)^{3/4}
    \biggl(\frac{L_\su{UV}/L_\su E}{0.1}\biggr)^{3/4}
    \biggr(\frac{\kappa_\su{IR}/\kappaT}{20}\biggr)^{-1/4}
    \biggr(\frac{\kappa_\su{UV}/\kappaT}{2000}\biggr)^{-1/4}\,
    \si{\solarmass\per\year}
  \end{equation}
}{
  \begin{multline}
  \num{\sim0.15}\,
    \biggl(\frac M{\SI{e7}{\solarmass}}\biggr)^{3/4}
    \biggl(\frac{L_\su{UV}/L_\su E}{0.1}\biggr)^{3/4}\times{} \\
  \biggr(\frac{\kappa_\su{IR}/\kappaT}{20}\biggr)^{-1/4}
    \biggr(\frac{\kappa_\su{UV}/\kappaT}{2000}\biggr)^{-1/4}\,
    \si{\solarmass\per\year}
  \end{multline}
}
and
\ifaastex{
  \begin{equation}
  \num{\sim1900}\,
    \biggl(\frac M{\SI{e7}{\solarmass}}\biggr)^{1/4}
    \biggl(\frac{L_\su{UV}/L_\su E}{0.1}\biggr)^{1/4}
    \biggr(\frac{\kappa_\su{IR}/\kappaT}{20}\biggr)^{1/4}
    \biggr(\frac{\kappa_\su{UV}/\kappaT}{2000}\biggr)^{1/4}\,
    \si{\kilo\meter\per\second}
  \end{equation}
}{
  \begin{multline}
  \num{\sim1900}\,
    \biggl(\frac M{\SI{e7}{\solarmass}}\biggr)^{1/4}
    \biggl(\frac{L_\su{UV}/L_\su E}{0.1}\biggr)^{1/4}\times{} \\
  \biggr(\frac{\kappa_\su{IR}/\kappaT}{20}\biggr)^{1/4}
    \biggr(\frac{\kappa_\su{UV}/\kappaT}{2000}\biggr)^{1/4}\,
    \si{\kilo\meter\per\second}
  \end{multline}
}
respectively. These equations have identical scaling as Equations~(34) and~(35)
in \citetalias{2016ApJ...825...67C}, and the pre-factors are similar to their
counterparts in that article, to wit, \SI{\sim0.1}{\solarmass\per\year} and
\SI{\sim5000}{\kilo\meter\per\second}. The temporal fluctuations of the two
quantities in our current simulations are by \SI{\approx11}{\percent} and
\SI{\approx26}{\percent} respectively.

For the \textquote{steady} state of the \irrmhd{} stage, we remove the
contribution due to \ac{UV} radiation by subtracting from each loss rate its
average value in the \uvrmhd{} stage. The mass loss rate and outflow speed
derived from these \ac{IR}-only loss rates are then scaled by \cref{eq:IR mass
loss rate,eq:IR outflow speed}. We obtain
\ifaastex{
  \begin{equation}
  \num{\sim0.7}\,
    \biggl(\frac M{\SI{e7}{\solarmass}}\biggr)^{3/4}
    \biggl(\frac{L_\su{UV}/L_\su E}{0.1}\biggr)^{3/4}
    \biggl[\frac{C_\su{IR}/(1-C_\su{IR})}{2.56}\biggr]^{3/4}
    \biggr(\frac{\kappa_\su{IR}/\kappaT}{20}\biggr)^{-3/4}
    \biggr(\frac{\kappa_\su{UV}/\kappaT}{2000}\biggr)^{1/4}\,
    \si{\solarmass\per\year}
  \end{equation}
}{
  \begin{multline}
  \num{\sim0.7}\,
    \biggl(\frac M{\SI{e7}{\solarmass}}\biggr)^{3/4}
    \biggl(\frac{L_\su{UV}/L_\su E}{0.1}\biggr)^{3/4}
    \biggl[\frac{C_\su{IR}/(1-C_\su{IR})}{2.56}\biggr]^{3/4}\times{} \\
  \biggr(\frac{\kappa_\su{IR}/\kappaT}{20}\biggr)^{-3/4}
    \biggr(\frac{\kappa_\su{UV}/\kappaT}{2000}\biggr)^{1/4}\,
    \si{\solarmass\per\year}
  \end{multline}
}
and
\ifaastex{
  \begin{equation}
  \num{\sim600}\,
    \biggl(\frac M{\SI{e7}{\solarmass}}\biggr)^{1/4}
    \biggl(\frac{L_\su{UV}/L_\su E}{0.1}\biggr)^{1/4}
    \biggl[\frac{C_\su{IR}/(1-C_\su{IR})}{2.56}\biggr]^{1/4}
    \biggr(\frac{\kappa_\su{IR}/\kappaT}{20}\biggr)^{3/4}
    \biggr(\frac{\kappa_\su{UV}/\kappaT}{2000}\biggr)^{-1/4}\,
    \si{\kilo\meter\per\second},
  \end{equation}
}{
  \begin{multline}
  \num{\sim600}\,
    \biggl(\frac M{\SI{e7}{\solarmass}}\biggr)^{1/4}
    \biggl(\frac{L_\su{UV}/L_\su E}{0.1}\biggr)^{1/4}
    \biggl[\frac{C_\su{IR}/(1-C_\su{IR})}{2.56}\biggr]^{1/4}\times{} \\
  \biggr(\frac{\kappa_\su{IR}/\kappaT}{20}\biggr)^{3/4}
    \biggr(\frac{\kappa_\su{UV}/\kappaT}{2000}\biggr)^{-1/4}\,
    \si{\kilo\meter\per\second},
  \end{multline}
}
with fluctuations by \SI{\approx21}{\percent} and \SI{\approx16}{\percent}
respectively.

The \ac{IR}-driven outflow has a greater mass loss rate but a smaller outflow
speed than the \ac{UV}-driven outflow; this is because the former has higher
density and occupies a larger solid angle (\cref{sec:outflow}). Both \ac{IR}-
and \ac{UV}-driven outflows fall within the observed ranges of mass loss rates
and outflow speeds of \ac{UV} and \xray{} warm absorbers, which are
respectively \SIrange{\sim e-4}{\sim10}{\solarmass\per\year}
\citep[e.g.,][]{1969ApJ...158..859A, 1999ApJ...516..750C, 2000A&A...354L..83K,
2000ApJ...535L..17K} and \SIrange{\sim100}{\sim2000}{\kilo\meter\per\second}
\citep[e.g.,][]{2005A&A...431..111B, 2011MNRAS.410.2274Z, 2012ApJ...753...75C}.

In view of the torus not being in a formal steady state (\cref{sec:steady
state}), it is remarkable that the mass loss rate and outflow speed of the
combined \ac{IR}- and \ac{UV}-driven outflow in the \textquote{steady} state of
the \irrmhd{} stage fluctuate only by \SI{\approx11}{\percent} and
\SI{\approx9}{\percent}. This constancy suggests that the properties of the
outflow may be determined solely by the conditions at the inner edge, which
does remain at approximately the same location throughout the
\textquote{steady} state.

\subsubsection{Morphology}
\label{sec:realistic morphology}

The degree to which gas pressure lends support against gravity is
$\mathrelp\sim(c_\su s/v_\phi)^2$, which according to \cref{eq:mass and sound
speed} is $\mathrelp\propto M^{-1/2}$. In contrast, the typical \ac{IR}
radiation energy density in the torus is always $\mathrelp\sim\aSB
T_\su{ds}^4$; \ac{IR} radiation pushes on the gas as it escapes the torus,
resulting in a typical acceleration of $\mathrelp\sim\aSB T_\su{ds}^4/(\rho
r_\su{ds})$, which is independent of $M$ insofar as $\rho r_\su{ds}$ is
constant. In other words, as $M$ increases, gas pressure support weakens while
\ac{IR} radiation pressure support remains roughly the same.

Vertical support against gravity in the \textquote{steady} state of the
\irrmhd{} stage is due to different forces in different places: \ac{UV}
radiation powers the wind; \ac{IR} radiation accelerates the wings
(\cref{sec:outflow}); gas and \ac{IR} radiation pressures contribute equally to
inflate the head (\cref{sec:forces}); the same pressures prop up the body, but
gas pressure contributes roughly twice as much as \ac{IR} radiation pressure
(\cref{sec:forces}). Based on the argument in the previous paragraph, we expect
when $M$ assumes realistic values, the wind and the wings would remain equally
well supported against gravity, hence neither their dimensions nor their
density fluctuations (\cref{sec:obscuration}) would change substantially in
character. The head and the body would lose respectively
$\mathrelp\approx\tfrac12$ and $\mathrelp\approx\tfrac23$ of their vertical
support and become geometrically thinner. Because gas and radiation influence
the distribution of each other, there may not be a strict proportionality
between $c_\su s/v_\phi$ and the aspect ratio of either head or body, but the
greater importance of \ac{IR} radiation in the head suggests that its height
should scale more slowly with $c_\su s/v_\phi$ than that of the body.

The latitude range of the wind at $\tau_\su{UV}\lesssim1$ is likely governed by
$\kappa_\su{UV}$; expressly, a higher $\kappa_\su{UV}$ implies a more slender
wind along the inner surface \citepalias{2016ApJ...825...67C}.

The properties of all four structures may vary with the typical torus density,
which is reflected in $\tau_\su T$. The density of the wind and the wings
should be independent of density elsewhere; this is because the rate of
momentum delivery to the torus by \ac{UV} radiation, the characteristic size of
the outflow, and the outflow speeds as given by \cref{eq:UV outflow speed,eq:IR
outflow speed} do not depend on density. The defining attribute of the head is
its vertical thickness relative to the body (\cref{sec:steady state}),
maintained in part by \ac{IR} and \ac{UV} radiation (\cref{sec:forces}); if the
torus is denser, the penetration depth of radiation is reduced, so the head may
appear less radially extended. The situation with the body is less certain due
to positive feedback: An increase in density makes the body less permeable to
\ac{IR} radiation, weakening not only \ac{IR} radiative support, but also gas
support since gas temperature is tied to \ac{IR} temperature
(\cref{sec:temperature}); the torus becomes thinner and even denser as a result
\citep[see also][]{2012ApJ...759...36R}. The feedback loop may be limited by
other physical effects that support the body vertically, such as Compton
heating (\cref{sec:Compton heating}). Contrarily, an initial decrease in
density triggers the feedback loop to run the opposite way, which causes the
body to become less and less dense.

\subsubsection{Temperature}
\label{sec:realistic temperature}

Energy balance and temperature in realistic tori are controlled by external
illumination; therefore, we expect \ac{IR} radiation to diffuse outward from
the inner edge, and temperature contours near the mid-plane to be not far from
vertical, no matter what $c_\su s/v_\phi$ is. In contrast, if internal
dissipation at the mid-plane were to dominate, \ac{IR} radiation would diffuse
from the mid-plane, which means temperature contours would make a sharp angle
at the mid-plane. Indeed, temperature contours in our simulation are quite
vertical far above the head and the body (\cref{sec:temperature}). We should
not interpret this as a sign that temperature contours are strictly vertical at
any $c_\su s/v_\phi$, but as a suggestion that their shape enjoys relative
independence from that of the head and the body, so they likely remain
vertically extended even as $c_\su s/v_\phi$ is reduced.

Assuming such, we could ask how the observed face-on and edge-on temperature
profiles (\cref{sec:photosphere}) change as a function of $c_\su s/v_\phi$. The
face-on \ac{IR} photosphere follows the outline of the head and the body. With
smaller $c_\su s/v_\phi$, both structures would be flatter; by virtue of the
verticality of temperature contours near the mid-plane, a radially outward
temperature gradient should always be observed. As for the temperature profile
of the edge-on \ac{IR} photosphere, the jump at large altitudes is a result of
density in the wings diminishing with height; as long as wings are
geometrically thick, the jump should remain.

\subsubsection{Obscuration}
\label{sec:realistic obscuration}

The obscuration properties of the torus are a direct consequence of its density
distribution. The soft \xray, \ac{IR}, and \ac{UV} covering fractions should
not vary strongly with $c_\su s/v_\phi$ or $\tau_\su T$; this is because the
primary obscurers are the wings and the wind (\cref{sec:obscuration}), which
should remain geometrically thick at lower $c_\su s/v_\phi$ and equally dense
at any $\tau_\su T$ (\cref{sec:realistic morphology}). However, the
Compton-thick fraction would decline with smaller $c_\su s/v_\phi$ because the
obscurers here are the head and the body (\cref{sec:obscuration}), both of
which would be thinner at reduced $c_\su s/v_\phi$ (\cref{sec:realistic
morphology}).

Another important aspect of \ac{AGN} obscuration is the distribution of
observed column densities. Stated in terms of simulation variables, the
observed $\ods\Omega{\log_{10}\tau}$ is generally flat for $\tau_\su
T\gtrsim0.01$, with a slight rise toward higher $\tau_\su T$ \citep[e.g.,][and
references therein]{2007A&A...463...79G}. Our torus already has flat
$\ods\Omega{\log_{10}\tau}$ for $\tau\lesssim0.4$ (\cref{sec:obscuration}), and
we get an even better agreement with observations if we consider how
$\ods\Omega{\log_{10}\tau}$ would change under smaller $c_\su s/v_\phi$ or
different $\tau_\su T$.

When $c_\su s/v_\phi$ is lowered, gas would gather toward the mid-plane, and
both head and body would take up a smaller solid angle around the origin; as a
consequence, the slender peaks in the right panel of \cref{fig:optical depths},
which reflect the densest parts of the torus, would move to the right, and the
area under them would diminish. There may still be the same gentle roll-off to
a plateau on the left side due to the wings and the wind, but the plateau would
be higher because the total area under the histogram is conserved. The
histogram would therefore become flatter overall, which means
$\ods\Omega{\log_{10}\tau}$ would be practically constant over a large span of
$\log_{10}\tau$.

Alternatively, realistic tori may have a wide range of $\tau_\su T$ extending
up to a few. Since our torus already has $\tau_\su T\sim2$ (\cref{sec:IR-RMHD
stage}), we are mostly interested in cases of smaller $\tau_\su T$, which have
left-shifted peaks. When $\ods\Omega{\log_{10}\tau}$ histograms for tori of
different $\tau_\su T$ are stacked, offset peaks would add up to form a flatter
distribution; in other words, the near constancy of $\ods\Omega{\log_{10}\tau}$
over $\log_{10}\tau$ could be a natural consequence of realistic tori having a
broad distribution of $\tau_\su T$.

\subsubsection{Photoionization}
\label{sec:photoionization}

Our simulation is best at treating \ac{IR} \ac{RT} at $\tau_\su{IR}\gtrsim1$,
or where the column density from the central source is
\SI{\gtrsim8e22}{\per\square\centi\meter}; it is less reliable inside the
central hole since we omit \ac{UV} and \xray{} photoionization. Here we
estimate how reinstituting these effects may modify our torus.

Photoionization raises gas temperature in the central hole
\citep{1986ApJ...308L..55K, 2001ApJ...561..684K}, and the higher gas pressure
could compress the torus \citep{2012ApJ...761...70D, 2016ApJ...819..115D}.
Although our simulation does not explicitly constrain pressure from
photoionized gas in the central hole, we can estimate its effect by computing
the ionization parameter
\begin{equation}
\Xi\eqdef\frac{L_\su{ion}}{4\pi r^2cp_\su t}
  =\frac{L_\su{ion}}{L_\su{UV}}\frac{\kappa_\su{IR}}{\kappa_\su{UV}}
  \biggl(\frac{p_\su t}{\aSB T_\su{ds}^4}\biggr)^{-1}
  \biggl(\frac r{r_\su{ds}}\biggr)^{-2},
\end{equation}
where $L_\su{ion}$ is the ionizing luminosity, and $p_\su t$ is the sum of gas
and \ac{IR} radiation pressure \citep{1981ApJ...249..422K}. \Ac{AGN}
photoionization calculations indicate that $\Xi$ locks in at \num{\sim10} if,
as here, a cool gas reservoir is present \citep{2001ApJ...561..684K}. Taking
$L_\su{ion}\sim L_\su{UV}$, we find that $\Xi$ in the head, body, wings, and
wind are \num{\sim1}, \num{\sim2}, \num{\sim2.5}, and \num{\sim1.2}
respectively. This means our torus is too pressurized to be confined by
photoionized gas; rather, its geometrical thickness is limited primarily by
gravity.

Hotter gas can also destroy dust grains through sputtering
\citep{1979ApJ...231...77D, 1994ApJ...431..321T}, thus reducing the
effectiveness of wind driving through radiation pressure on dust. The
sputtering timescale at gas temperature of \SI{\sim e5}{\kelvin} is
$\num{\sim100}\,[n_\element
H/(\SI{e6}{\per\cubic\centi\meter})]^{-1}[a/(\SI{0.1}{\micro\meter})]\,\si{\year}$,
where $n_\element H$ and $a$ are the number density of hydrogen atoms and the
radius of dust grains respectively \citep{1979ApJ...231...77D,
1994ApJ...431..321T}, and it decreases sharply at higher temperatures until
\SIrange{\sim e7}{\sim e7.5}{\kelvin}. This is not very different from the time
it takes for the \ac{UV}-driven wind to escape from the inner edge at
$R=R_\su{in}$ to infinity:
\ifaastex{
  \begin{equation}
  \frac{R_\su{in}}{v_\infty}\sim55\,
    \biggl(\frac M{\SI{e7}{\solarmass}}\biggr)^{1/4}
    \biggl(\frac{L_\su{UV}/L_\su E}{0.1}\biggr)^{1/4}
    \biggr(\frac{\kappa_\su{IR}/\kappaT}{20}\biggr)^{-3/4}
    \biggr(\frac{\kappa_\su{UV}/\kappaT}{2000}\biggr)^{1/4}
    \biggl(\frac{R_\su{in}}{r_\su{ds}}\biggr)^{3/2}\,\si{\year}.
  \end{equation}
}{
  \begin{multline}
  \frac{R_\su{in}}{v_\infty}\sim55\,
    \biggl(\frac M{\SI{e7}{\solarmass}}\biggr)^{1/4}
    \biggl(\frac{L_\su{UV}/L_\su E}{0.1}\biggr)^{1/4}\times{} \\
  \biggr(\frac{\kappa_\su{IR}/\kappaT}{20}\biggr)^{-3/4}
    \biggr(\frac{\kappa_\su{UV}/\kappaT}{2000}\biggr)^{1/4}
    \biggl(\frac{R_\su{in}}{r_\su{ds}}\biggr)^{3/2}\,\si{\year}.
  \end{multline}
}
The dust sublimation radius $r_\su{ds}$ is given by \cref{eq:dust sublimation
radius}, and the wind speed is defined in \citetalias{2016ApJ...825...67C}. The
dust content of the outflow therefore depends sensitively on how quickly cold
gas evaporated from the inner surface rises in temperature.

Photoionization could impart momentum to the wind through lines, similar to the
accretion disk wind of \citet{2000ApJ...543..686P} and the
radiation-accelerated magnetocentrifugal wind of \citet{2005ApJ...631..689E}.
Heating by photoionization can also produce a thermally driven wind
\citep{1986ApJ...308L..55K, 1993ApJ...402..109B, 2001ApJ...561..684K,
2005A&A...431..111B}. This suggests some kind of complementarity between
outflow-driving mechanisms: In places where $\Xi\lesssim1$, gas temperature is
\SI{\lesssim e4.5}{\kelvin} \citep{1981ApJ...249..422K}, so gas may stay dusty
long enough for radiation pressure to evacuate it from the \ac{AGN}; in places
where $\Xi\gtrsim1$, gas temperature is near Compton
\citep{1981ApJ...249..422K}, so gas may already have enough thermal pressure to
expel itself.

\subsubsection{Compton heating}
\label{sec:Compton heating}

We ignore volumetric Compton heating in our simulation, but \xrays{} at
luminosities a few times weaker than in the \ac{UV} can produce sufficient
\ac{IR} radiation deep inside the torus to induce important changes to the
distributions of density and specific angular momentum, particularly near the
mid-plane \citep{2008ApJ...679.1018S}. As a matter of fact, since the mid-plane
is optically thick in the \ac{IR} but only marginally so in \xrays, \xrays{}
could be the deciding factor in how concentrated gas is near the mid-plane. A
positive feedback loop could exist, wherein Compton heating injects the energy
requisite to seed a vertical \ac{IR} radiative flux, which lowers the density
near the mid-plane and allows \ac{IR} radiation from the inner edge to enter
the torus and support it \citep[see also][]{2012ApJ...759...36R}.

Compton heating most certainly changes gas and \ac{IR} temperatures at
$\tau_\su{UV}\gtrsim1$. If the torus is Compton thin in all directions and has
approximately the same density everywhere, we would expect Compton heating to
be uniform, hence temperature contours would be vertical near the mid-plane as
if \xrays{} were absent (\cref{sec:realistic temperature}). But if the torus
transitions from Compton thick to Compton thin with increasing latitude, then
Compton heating would occur predominantly near the mid-plane, at a rate
diminishing with distance from the central source and with effect resembling
internal dissipation (\cref{sec:realistic temperature}); furthermore, if
sufficient energy is deposited at the mid-plane that the secondary \ac{IR}
radiative flux has noticeable effect on vertical support, temperature contours
may no longer be vertical near the mid-plane.

\subsection{Relation of our model to literature}
\label{sec:comparison}

This article does not report the first attempt at understanding torus dynamics.
On the topic of outflows alone, several authors, we included, have written
extensively on how outflows could account for torus phenomenology. We situate
the present work relative to past ones by summarizing the contrasts.

\subsubsection{Comparison with our previous work}

The approaches taken in \citetalias{2016ApJ...825...67C} and here for the
\uvrmhd{} stage (\cref{sec:angular momentum}) are quite complementary. In the
earlier simulations, we fixed a sub-Keplerian $j(R)$ and searched for
$L_\su{UV}$ such that the torus is reasonably long-lived; we concluded that
$L_\su{UV}$ should be small to allow under-supported gas to flow toward the
inner edge, but not so small that the gas contracts into the dust sublimation
surface. In our current simulations, we choose $L_\su{UV}$ and determine $j(R)$
that best promotes torus longevity (\cref{sec:parameter study}); we discover
that the torus can survive longer if $j(R)$ is sub-Keplerian and gas feeds the
inner edge, but there is a limit to how far below Keplerian we can go before
the torus shrinks to smaller than the dust sublimation surface.

\subsubsection{Comparison with magnetocentrifugal wind models}

\Citet{1994ApJ...434..446K} proposed that obscuration in \acp{AGN} could be
provided by a magnetocentrifugal wind, which, intuitively speaking, is a
centrifugally driven outflow guided by open magnetic field lines. Their model
presupposes a razor-thin accretion disk as the mass source for the wind; our
simulation allows a given amount of gas to evolve without prescribing how mass
resupply occurs. Their model also requires the specification of several
parameters for the wind, among which are the launch radius from the accretion
disk, the spherically radial density profile, and the conserved mass flux and
angular momentum along the streamline, whereas our simulation fixes only the
initial angular momentum profile (\cref{sec:angular momentum}). Most
importantly, their model posits the existence of a large-scale, dynamically
dominant magnetic field; in contrast, we assume only that the \ac{MRI}
amplifies magnetic field contained entirely within the gas. We further find
that radiation alone can lift gas from the inner edge into a high-latitude
outflow (\cref{sec:outflow}). Despite our torus being magnetized, this outflow
is not a magnetocentrifugal wind because meandering loops of magnetic field in
the outflow are too weak to exert much force; instead, they are passively
dragged out by gas motion (\cref{sec:magnetic field}).

\Citet{1994ApJ...434..446K} acknowledged the importance of \ac{IR} and \ac{UV}
radiation pressure on dust, but \citet{2005ApJ...631..689E} was the first to
study how \iac{AGN}-like spectrum photoionizes a magnetocentrifugal wind and
injects momentum through atomic absorption. He put shielding gas of arbitrary
column density into the model to prevent overionization of the wind, but
omitted dust, which could also influence photoionization by removing \ac{UV}
radiation. \Citet{2012ApJ...749...32K} expanded the work of
\citet{2005ApJ...631..689E} by adding dust opacity and momentum transfer from
dust absorption. Neither model considers pressure on dust from reprocessed
\ac{IR} radiation, which transmits momentum and energy deposited by \ac{UV}
radiation through the torus, and which our simulation identifies as critical
for maintaining an outflow at $\tau_\su{UV}\gtrsim1$ (\cref{sec:outflow}).

\Citetseealso{2006ApJ...648L.101E}{}{1999ApJ...513..180K} examined another
variation on the magnetocentrifugal wind, one in which dusty gas is clumped;
the authors assumed clumps are individual entities and touched on how they may
be magnetically confined. Our simulation sheds light on both issues. The first
and second panels of the bottom row of \cref{fig:overview} depict the
inhomogeneous density distribution in the wings and the wind, which obscure at
high latitudes (\cref{sec:obscuration}). We find wedges of various densities in
the wind, density ridges in the wings, and even a hook-shaped feature near the
top of the second panel. The density profiles in \cref{fig:density} illustrate
the same point in a different way. Our simulation therefore underlines the
point that the common picture of spherical, well-separated clumps must not be
taken too seriously. Furthermore, density perturbations in our torus are not
static or stationary structures confined externally by gas or magnetic
pressure, or internally by self-gravity; they are imprinted in the wind by
bursty wind launching (\cref{sec:steady state}), and in the wings by bursts of
newly launched, faster gas shocking with slower gas further out (\cref{sec:UV
dynamics}). Density perturbations are ephemeral; only by virtue of their
frequent recurrence at the same place with the same morphologies do they become
consistent features of the torus.

\subsubsection{Comparison with photoionization-driven models}

\Citetseealso{2012ApJ...758...66W}{2015ApJ...812...82W}{2014MNRAS.445.3878S,
2016ApJ...828L..19W} explored through simulations the idea of a gas fountain
powered by the central source through \ac{UV} radiative acceleration, \xray{}
photoionization heating, and Compton heating. The neglect of heating by \ac{UV}
radiation from the central source in these simulations precludes the treatment
of thermal \ac{IR} radiation. When gas in these simulations moves out of the
central hole, it receives reduced radiative acceleration and falls back to the
mid-plane; this could be because these simulations presume \ac{UV} radiation is
concentrated in the polar direction, and because they ignore \ac{IR} radiation,
which transports momentum and energy to $\tau_\su{UV}\gtrsim1$ and thus extends
the outflow into that region (\cref{sec:outflow}). \Citet{2016ApJ...828L..19W}
combined in one simulation turbulence generated by supernovae
\citep{2002ApJ...566L..21W} with driving by the central source, but stars
cannot make gas geometrically thick on parsec scales
\citep{1988ApJ...329..702K}.

\subsubsection{Comparison with other models with \texorpdfstring{\acs*{IR}}{IR}
radiation}

\Citeauthor{2012ApJ...761...70D} conducted a series of simulations to
investigate whether \ac{IR} radiation pressure on dust can create a
geometrically thick torus. Our simulation most resembles those by
\citet{2012ApJ...747....8D}: Their simulations have \ac{IR} radiation driving a
wide-angle outflow, while ours have \ac{IR} and \ac{UV} working in concert to
achieve the same effect (\cref{sec:outflow}). However, there are also important
differences between the two sets of simulations. Their simulations postulate a
razor-thin accretion disk as a mass source for the outflow, the
characterization of which introduces additional free parameters not
self-consistently determined by the simulation; we avoid this by putting all
the mass in the simulation domain right at the start and letting it develop
structures on its own. The outflow in their simulations originates from across
the entire accretion disk because their mid-plane boundary condition assures
so; in our simulation, where no such boundary condition is assumed, gas is
launched into the outflow exclusively at the torus inner edge (\cref{sec:steady
state}). Finally, the \ac{IR}-driven outflow in their simulations is a failed
wind that apparently falls back to the mid-plane. Our simulation with both
\ac{IR} and \ac{UV} radiation tells a different story: The \ac{UV}-driven
outflow at $\tau_\su{UV}\lesssim1$ is gravitationally unbound, and at least
part of the \ac{IR}-driven outflow at $\tau_\su{UV}\gtrsim1$ is unbound
(\cref{sec:outflow}).

The later simulations by \ifaastex{\citet{2012ApJ...761...70D}
and~\citet{2016ApJ...819..115D}}{\citet{2012ApJ...761...70D,
2016ApJ...819..115D}} consider how the central source deposits momentum through
\ac{UV} radiation, and momentum and energy through \xrays; they diverge
markedly from the simulations by \citet{2012ApJ...747....8D} and from ours. In
both their simulations and ours, \ac{IR} radiation is created when the central
source heats dust; the difference is that in our simulation, heating through
\ac{UV} radiation is concentrated at the inner edge
\citepalias{2016ApJ...825...67C}, whereas in their simulations, heating is more
widespread because it is due to \xrays, not \ac{UV} radiation. In their
simulations, \ac{UV} radiation transfers momentum solely through lines; this
implies that the authors were looking at a different situation from ours,
namely, one with a dust-free central hole, but whether dust is present depends
on how fast photoionization raises gas temperature to Compton
(\cref{sec:photoionization}). Lastly, photoionization heating in their
simulations produces a hot atmosphere that envelopes and vertically squeezes
their tori; therefore, \ac{IR} radiation can only push an outflow along, not
far above, the mid-plane. Our torus, in comparison, has high enough gas and
\ac{IR} radiation pressure that photoionized gas is not the main determinant of
its geometrical thickness (\cref{sec:photoionization}).

The recent simulations by \citet{2016MNRAS.460..980N} study how dusty gas
interacts with \ac{UV} radiation and \xrays{} from the central source, and with
reprocessed \ac{IR} radiation; dust absorption, Compton recoil, and a variety
of chemical processes transfer energy and momentum between gas and radiation.
Their simulations always produce geometrically thin structures and thus do not
explain vertical support in tori. There are several reasons that might explain
this thinness. Their initial condition is geometrically thin and therefore has
Keplerian rotation, but Keplerian rotation is inconsistent with a long-lived,
geometrically thick structure (\cref{sec:angular momentum}). Their central
source radiates zero flux in the mid-plane, so little radiation enters the gas;
in addition, their geometrically thin structure is poorly resolved in the
vertical direction at small radii. Lastly, a photoionized and Compton-heated
atmosphere surrounds their geometrically thin structure and confines the cooler
gas to the mid-plane.

\subsubsection{Summary of comparisons}

In sum, our simulation offers a new perspective on torus dynamics. With the
bare minimum of physics, to wit, momentum and energy coupling between gas,
\ac{IR} radiation, and \ac{UV} radiation, our simulation demonstrates that
radiation on its own can propel an outflow far above the mid-plane; a
mass-loading mechanism and a strong magnetic field steering gas to high
latitudes, as in a magnetocentrifugal wind, are unnecessary
(\cref{sec:outflow}). Moreover, our simulation distinguishes itself from gas
fountain models by showing that \ac{IR} radiation is pivotal in delivering
momentum and energy to $\tau_\su{UV}\gtrsim1$, thereby driving a wide-angle
outflow (\cref{sec:outflow}). Although photoionization can augment the outflow
in the central hole (\cref{sec:photoionization}) and Compton heating can modify
the shape of structures deep inside the torus (\cref{sec:Compton heating}),
neither is likely to change the fundamental character of the radiation-driven
outflow.

Our simulations in \citetalias{2016ApJ...825...67C} and here also highlight two
aspects of tori that have not always received the attention they deserve.
First, efforts to understand the observed \ac{IR} spectrum of \acp{AGN} often
assume torus gas is in clumps, and these clumps are almost invariably taken to
be spherical, discrete, and pressure-confined
\citep[e.g.,][]{2006ApJ...648L.101E}. In contrast, our current simulation
suggests that fleeting, irregular density perturbations can arise simply from
cool gas accelerating to high speeds due to radiation pressure, then shocking
with slower gas (\cref{sec:outflow}). Detailed \ac{RT} calculations will be
needed to assess whether such density inhomogeneities produce strong enough
\ac{FIR} emission and a shallow enough silicate feature to match observations.
Second, our previous and current simulations both portray the torus as a
flow-through system: Gas is conveyed inward along the mid-plane; the majority
of this gas is expelled in the outflow, and only a small fraction is captured
by the central mass to fuel the generation of \ac{UV} radiation. This view has
already been taken by \citet{1988ApJ...329..702K}, but our work makes clear
that sub-Keplerian rotation is necessary for maintaining a steady-state torus
in the presence of strong radiation pressure (\cref{sec:angular momentum}), and
we have provided an estimate of the mass resupply rate requisite for steady
state (\cref{sec:realistic outflow}).

\section{Conclusions}

We have performed three-dimensional, time-dependent \acp{RMHD} simulations of
\ac{AGN} tori featuring quality \ac{RT} and simultaneous evolution of gas and
radiation. For the first time, our torus achieves a \textquote{steady} state
lasting for more than an orbit at the inner edge, and potentially for much
longer. This \textquote{steady} state is defined as the torus having constant
overall morphology (\cref{sec:steady state}). It is obtained by reducing the
angular momentum profile before the simulation starts (\cref{sec:angular
momentum}), which raises the total binding energy and thus allows the torus to
survive longer under \ac{UV} radiation doing positive work (\cref{sec:parameter
study}).

The existence of a \textquote{steady} state is significant: While tori in
previous simulations could not endure \ac{UV} irradiation for more than two
orbits at the inner edge \citepalias{2016ApJ...825...67C}, our current
simulation demonstrates that a torus with the right parameters can indeed
remain in a steady state for multiple orbits (\cref{sec:parameter study}).
Granted that our torus cannot formally reach equilibrium owing to our choice of
$\kappa_\su{UV}$ \citepalias{2016ApJ...825...67C} and to computational cost, we
can already learn much from its approximate \textquote{steady} state that would
conceivably carry over to the true steady state. Moreover, the ability to study
the torus in a quasi-stationary state boosts our confidence in separating the
\textquote{steady} state (\cref{sec:steady state}) from transitory behavior
(\cref{sec:transient}).

We perceive four \textquote{steady}-state structures in the torus, namely,
head, body, wings, and wind (\cref{sec:steady state}). Vertical support against
gravity is dominated by gas pressure in the head and the body, \ac{IR}
radiation pressure in the wings, and \ac{UV} radiation pressure in the wind
(\cref{sec:forces}). By inspecting the \textquote{steady}-state flow and the
forces driving it, we realize that these structures are not hydrostatic.
Instead, due to insufficient support against gravity, most gas falls toward the
inner edge through the body and the head; as gas reaches the inner edge, it
flies outward on \ac{UV} radiative acceleration (\cref{sec:streamlines}). The
outflow is initially directed at high latitudes, but it spreads out in solid
angle once it climbs above the head and the body (\cref{sec:outflow}). The part
remaining in the central hole is the wind; it is propelled by \ac{IR} and
\ac{UV} radiation. The part expanding beyond the central hole is the wings; it
is powered by \ac{IR} radiation (\cref{sec:outflow}). The four structures are
simply regions that hold on to their shapes as gas flows through them.

The study of forces clarifies the subtle role \ac{IR} radiation plays in torus
dynamics: It opens up the central hole (\cref{sec:transient}), partially
supports the body and the lower parts of the wings in the vertical direction
(\cref{sec:forces}), and drives an outflow in the wings where \ac{UV} radiative
acceleration fails (\cref{sec:outflow,sec:realistic outflow}).

It is reassuring that most statements pertaining to \ac{RHD} tori are valid
here as well: Gas and \ac{IR} radiation have equal temperature inside the
optically thick torus (\cref{sec:temperature}), the torus focuses \ac{IR}
radiation toward the axis (\cref{sec:IR}), and the outflow has mass loss rate
and speed consistent with observations (\cref{sec:realistic outflow}). The
strong resemblance between \ac{RHD} and \ac{RMHD} tori suggests that the effect
of magnetic field over timescales as short as a few orbits is small
(\cref{sec:magnetic field}). It bears reiterating that the influence of
magnetic field on realistic tori is felt only over many orbits, as \ac{MHD}
stresses redistribute angular momentum and thereby set the steady-state angular
momentum profile.

Observational predictions can be more easily made for a torus in
\textquote{steady} state. When seen face-on, the temperature profile of our
torus should follow the radially outward temperature gradient of the body,
which is $T\propto R^{-0.73}$ in our simulation; when seen edge-on, a jump in
temperature should be seen at high altitudes (\cref{sec:photosphere}). The
tenuous wings and wind obscure the central source in soft \xrays, the \ac{IR},
and the \ac{UV}, while the dense head and body also stop hard \xrays. The soft
\xray, \ac{IR}, and \ac{UV} covering fractions are all approximately three
quarters, which is close to the observed fraction of type\nobreakdash-2
\acp{AGN} (\cref{sec:obscuration}). Furthermore, if we assume \acp{AGN} have a
finite range of mid-plane column densities, then our torus also naturally
explains why the distribution of observed \ac{AGN} gas columns over logarithmic
column density is flat (\cref{sec:realistic obscuration}).

The torus around a given central mass $M$ is governed by three important
parameters: the Eddington ratio $L_\su{UV}/L_\su E$ of the central source, the
Thomson optical depth $\tau_\su T$ of the torus, and the angular momentum
profile $j(R)$ of the same \citepalias{2016ApJ...825...67C}. The first two
parameters are fixed in our simulations by observational constraints
(\cref{sec:angular momentum,sec:parameters}); the only freedom we have is with
$j(R)$ (\cref{sec:angular momentum}). Yet, with practically no fine-tuning, our
torus naturally arrives at a \textquote{steady} state typified by a
high-latitude, wide-angle outflow whose obscuration properties agree fairly
well with observations. Such outflow therefore deserves serious consideration
as a model for geometrically thick obscuration in \acp{AGN}.

\vskip\bigskipamount\noindent
The authors thank the anonymous referee for constructive comments. They are
grateful to Jim Stone, Yanfei Jiang, and Shane Davis for generously allowing
Athena and its time-dependent \ac{RT} module to be used for this project. This
research was partially supported by NASA/ATP grant NNX14AB43G and NSF grant
AST-1516299. C.H.C. acknowledges support from an ISF--CNSF grant, ERC advanced
grant \textquote{TReX}, and ISF I-CORE \textquote{Origins}. The simulations
were performed on the Johns Hopkins Homewood High-Performance Cluster and the
Maryland Advanced Research Computing Center.

\ifaastex{\bibliography{torus}}{\printbibliography}

\end{document}